 \documentclass[final,3p,times,twocolumn,]{elsarticle}

\usepackage{lineno,hyperref}
\modulolinenumbers[5]

\usepackage{color}
\usepackage{amssymb}
 \usepackage{amsmath}
 \usepackage{float}
\begin{document}

\begin{frontmatter}

\title{Autofocusing and self-healing of partially blocked circular Airy derivative beams}

\author{Anita Kumari}
\author{Vasu Dev}
\author{Vishwa Pal}
\ead{vishwa.pal@iitrpr.ac.in}

\address{Department of Physics, Indian Institute of Technology Ropar, Rupnagar 140001, Punjab, India}

\begin{abstract}
 We numerically and experimentally study the autofocusing and self-healing of partially blocked circular Airy derivative beams (CADBs). The CADB consists of multiple rings, and partial blocking of CADB with different kinds is achieved by using symmetric and asymmetric binary amplitude masks, enabling blocking of inner/outer rings and sectorially. The CADB blocked with different types possesses the ability to autofocus, however, the required propagation distance for abrupt autofocusing vary with the amount and types of blocking. The abrupt autofocusing is quantified by a maximum k-value, and how fast it changes around the autofocusing distance ($z_{af}$). In particular, CADB blocked with inner rings (first/two/three) exhibits an abrupt autofocusing, as the k-value sharply increases [decreases] just before [after] $z_{af}$. The maximum k-value always occurs at $z_{af}$, which decreases as the number of blocked inner rings increases. For CADB blocked with outer rings, the k-value gradually changes around $z_{af}$, indicating a lack of abrupt autofocusing. The value of $z_{af}$ increases with the number of blocked outer rings. This suggests that although outer rings contain low intensities, these play an important role in autofocusing. A sectorially blocked CADB possesses an abrupt autofocusing, and maximum k-value depends on the amount of blocking. The CADB blocked with different types possesses good self-healing abilities, where blocked parts reappear as a result of redistribution of intensity. The maximum self-healing occurs at $z_{af}$, where an overlap integral approaches a maximum value. Finally, we have compared ideal CADB and partially blocked CADB having the same radii, and found that an ideal CADB possesses better abrupt autofocusing. We have found a good agreement between the numerical simulations and experimental results. 
\end{abstract}


\end{frontmatter}

%\linenumbers

\section{Introduction}
\label{intro}
Optical beams possessing propagation properties such as abrupt autofocusing, self-acceleration, and self-healing are desired for various applications, such as in optical communications, optical tweezers, particle trapping and material processing \cite{ren2021}. Airy beam has made a considerable interest over the last several years, as it possesses a unique feature of moving along a parabolic trajectory (self-accelerating) without any optical elements, and good self-healing abilities against the obstructions \cite{efremidis2019}. These beams with the radial symmetry are known as circular Airy beams (CABs), and are investigated due to their abrupt autofocusing, where intensity increases abruptly by the several orders of magnitude just before the focal point (autofocusing distance) \cite{efremidis2010,chremmos2011,chremmos2012}. The abrupt autofocusing of CABs makes it an ideal candidate for several applications such as in biomedical treatment, high-resolution imaging, micro-engineering with lasers, and for trapping particles at different positions \cite{zhang2011,jiang2013}. There have been several studies on the CABs, such as CAB shaping by annular arrayed-core fiber, CAB with optical vortices, abrupt autofocusing of blocked CAB, propagation characteristics of partially coherent CAB, controllable autofocusing of conical CAB, propagation of sharply autofocused ring Airy Gaussian vortex beams, CAB with chirped factors, elliptical Airy beams, dual autofocusing of CAB with different initial launch angles, trapping of particles by modified CAB, vortex carrying CAB in free space and in turbulent media, generating arbitrary ways of circular Airy Gaussian vortex beams with single digital hologram, self-healing of focused CAB and radially Airy beams \cite{deng2019,jiang2012,li2014,jiang2018,zhang2017,zha2018,wang2022,jiang2016,lazer2023,wang2021,chen2020,zhang2022}.

Recently, circular Airy derivative beams (CADBs) have been investigated that include the $n^{th}$ order derivative of Airy function, which provides stronger abrupt autofocusing than that of CAB with similar beam parameters \cite{zang2022}. The propagation properties of CADB, chirped factors and array of CADB have also been studied for improving their strong abrupt autofocusing properties \cite{zang2022propagation,zang2022effect,zang2023,yu2023,zang2022dependence}. In many applications optical beams require to propagate through various components as well as disorder media (e.g., turbulent and biological tissues), which may cause the obstructions of the beam, and due to that the information carried by the beam can be lost. If the beam possesses self-healing abilities, the loss of the information can be prevented. The self-healing of optical fields refers to their ability to reconstruct themselves after a partial obstruction placed in their propagation path, which is one of the most interesting property that makes such beams very useful. There are several types of beams that possess self-healing property, which includes Bessel beam, cosine beam, polycyclic tornado swallow tail circular beam, aberration laser beam, Airy beam and Helico conical beam \cite{chu2012,bencheikh2020,zhang2022experimental,dev2021,hermosa2013,broky2008}. 

The self-healing of CABs has been shown to depend on the size, shape and position of blocking \cite{chen2020}. Further, it has been shown that the autofocusing distance of blocked CAB remains the same, and its autofocusing is found to be enhanced by a factor in a range between $[2.7-3.8]$, when a few first inner rings in the beam are blocked \cite{li2014}. As mentioned above, the CADBs possess stronger abrupt autofocusing than that of a normal CAB, and focused intensity becomes higher by the several factors. Thus, CADBs are found to be more suitable for the applications. So far, the propagation properties of obstructed CADBs have not yet explored. In particular, how the partially blocked CADB autofocuses under different conditions? And, how does it self-heal under various types of blocking? 

Here, we have investigated the propagation properties (autofocusing and self-healing) of CADBs, which are partially blocked symmetrically and asymmetrically at the initial plane ($z=0$). A detailed quantification of these properties of partially blocked CADBs is performed. In section 2, we have given a theoretical description of CADB. Section 3 consists of experimental generation of CADB and partially blocked CADB, the numerical simulations as well as their detailed quantification. In section 4, we provide the comparison of CADB and blocked CADB having the beam parameters. Finally, in section 5 concluding remarks are presented.
\section{Theoretical description}
\label{theo_desc}
The electrical field of a CADB propagating along the $z-$axis in an initial plane ($z=0$) is given by \cite{zang2022}
\begin{equation} 
 E(r,\phi,0)=\ Ai^{(n)}\Big(\frac{r_0 - r}{w_0}\Big) \exp\Big(a\frac{r_0 - r}{w_0}\Big), ~~~~~~r\leq R\label{eq1}
\end{equation}
where $r=\sqrt{(x^2+y^2)}$ and $\phi=\tan^{-1}y/x$ represent the radial distance and azimuthal angle, respectively. $a$ is an exponential decay factor, $\ w_0$ is a scaling factor, $\ r_0$ is the radius of CADB at $z=0$ plane and $R$ denotes the mean screen radius. $\ Ai^{(n)}(^.)$ is $n^{th}$-order derivative of the Airy function with respect to $r$. For $n=0$, Eq.\,(\ref{eq1}) converts to an expression of CAB. The CADBs possess autofocusing property, which is quantified by intensity contrast called the k-value. The k-value is defined as $I_{\mathrm{max}}(z>0)/I_{\mathrm{max}}(z=0)$, where $I_{\mathrm{max}}(z>0)$ and $I_{\mathrm{max}}(z=0)$ are the maximum intensities observed at $z>0$ and $z=0$ planes, respectively \cite{li2014}. It has been shown that the abrupt autofocusing increases with the order $n$ \cite{zang2022}. For $n=1$, the generalized CADB is also known as circular Airy prime beam (CAPB) \cite{zang2022}. In the present study, we have investigated the propagation properties of partially blocked CADBs with the order $n=1$. The beam parameters are chosen to be $r_0=1$ mm, $w_0=0.1$ mm, $\lambda=1064$ nm, and $a=0.1$ throughout the manuscript. 

\begin{figure}[!ht]
\centering
\includegraphics[height = 3.25 cm, keepaspectratio = true]{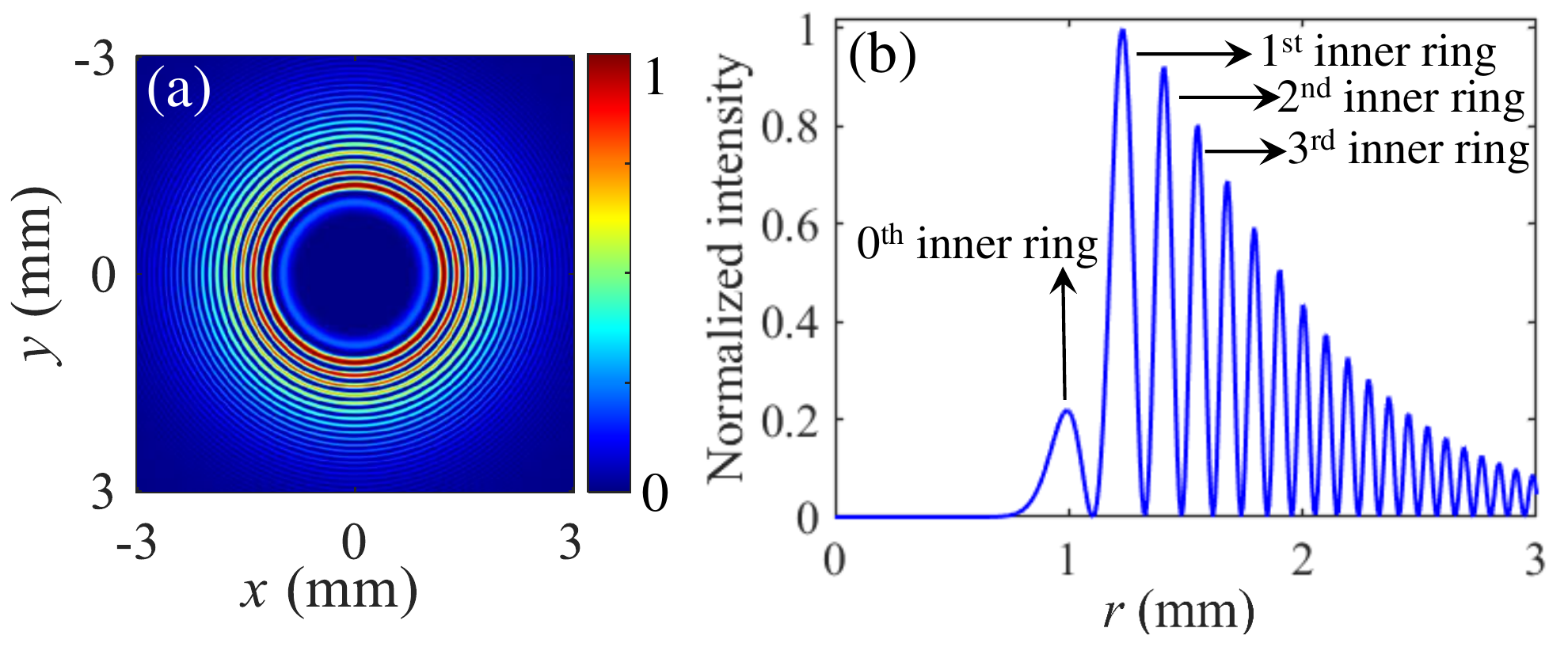}
\caption{(a) Normalized intensity distribution of CADB at the initial plane $z=0$. (b) Intensity cross section taken along the radial direction in (a). From the center ($r=0$), the inner rings are marked as $0^{\mathrm{th}}$, $1^{\mathrm{st}}$, $2^{\mathrm{nd}}$ and $3^{\mathrm{rd}}$, respectively.}
\label{fig1}
\end{figure}

The intensity distribution of CADB (Eq.\,(\ref{eq1})) is shown Fig.\,\ref{fig1}.
As evident, the intensity is distributed in multiple rings, and it decreases along the radial direction (away from the center). We have denoted from the center ($r=0$), an inner ring with low intensity as $0^{\mathrm{th}}$, and the most intense inner ring as $1^{\mathrm{st}}$ and subsequently inner rings with decreasing intensities as $2^{\mathrm{nd}}$, $3^{\mathrm{rd}}$..., as shown in Fig.\,\ref{fig1}(b). The radius and peak intensity of $0^{\mathrm{th}}$ ring are $0.991$ mm and $0.21$, respectively, showing that it consists of very small intensity. The $1^{\mathrm{st}}$, $2^{\mathrm{nd}}$, and $3^{\mathrm{rd}}$ inner rings have the radii $1.231$ mm, $1.406$ mm and $1.552$ mm, and their peak intensities are $1$, $0.92$ and $0.80$, respectively. Note, the maximum value of intensity is normalized to 1. As the CADB propagates along the axial direction ($z$-axis), due to lateral self-acceleration the outermost rings start bending in a curved manner and as a result of interference with adjacent rings the number of rings decreases gradually. Finally, after propagating a certain distance, the intensity becomes tightly focused in a narrow region (known as autofocusing distance), and the k-value is found to be $\sim357$, which is $\sim 7$ times higher than that of a CAB (having the same parameters) \cite{ zang2022}. 

An expression of autofocusing distance for a common CAB can be given as \cite{li2014, zang2022}
\begin{equation} 
 z_{af}=\frac{4\pi w_0}{\lambda}\sqrt{R_0 w_0}, \label{eq2}
\end{equation}
where $R_0$ is the radius of the first innermost intense ring. It has been shown that the same expression can be used for calculating an approximate autofocusing distance of CADB \cite{zang2022}. 
 
\begin{figure}[!ht]
\centering
\includegraphics[height =5.3cm, keepaspectratio = true]{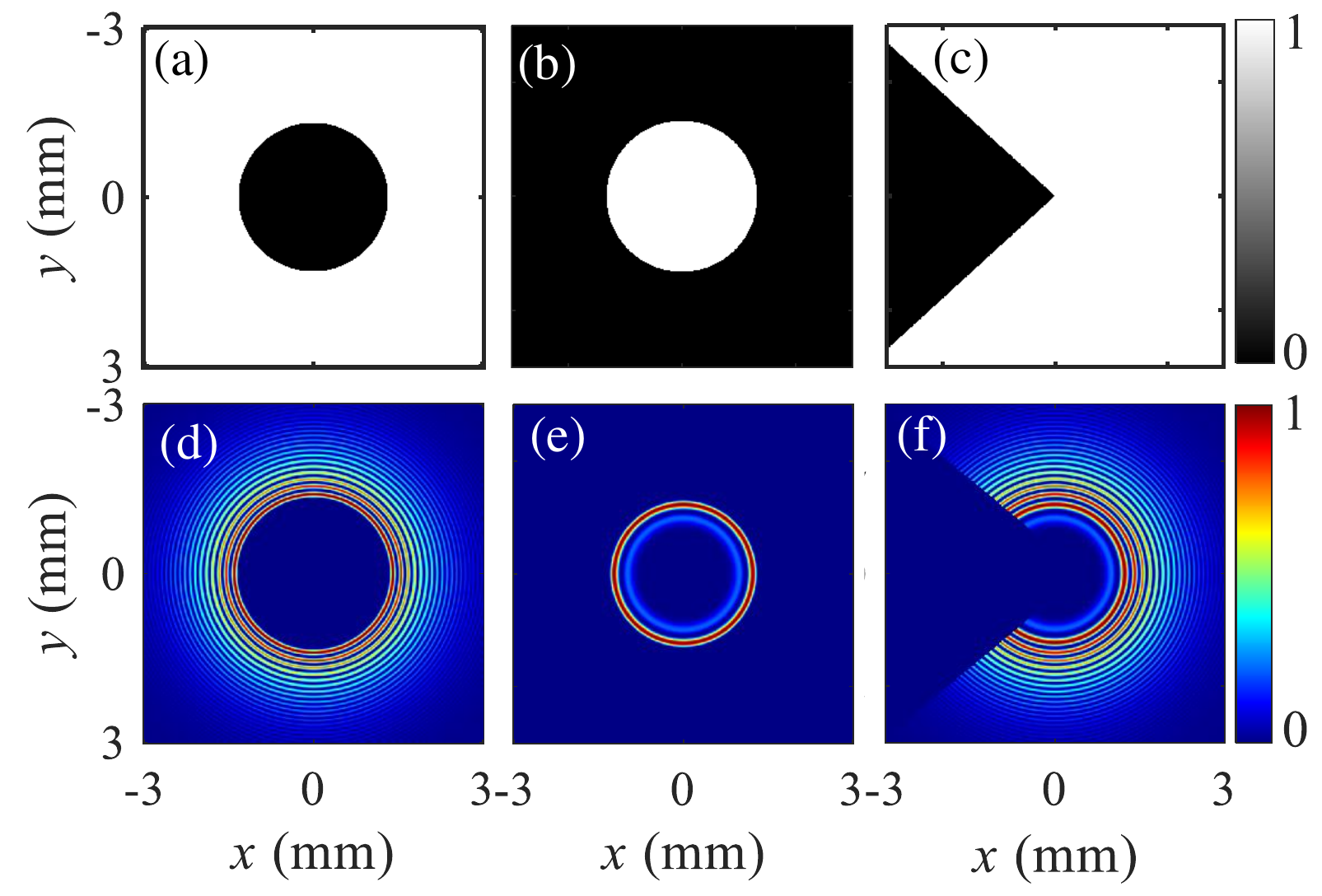}
\caption{(a)-(c) Different apertures with binary amplitudes for blocking regions from inside, outside and sectorial (Eqs.\,(\ref{eq3}) and (\ref{eq4})). The intensity distribution of blocked CADBs at initial plane (z=0) when (d) $0^{\mathrm{th}}$ $\&$ $1^{\mathrm{st}}$ inner rings are blocked, (e) outer rings are blocked, and (f) sectorially blocked.}
\label{fig2}
\end{figure}

In the present work, we have investigated the autofocusing and self-healing of CADBs partially blocked symmetrically and asymmetrically. The symmetric blocking of CADB is done with two types: i) from the center ($r=0$), blocking of inner rings, e.g., up to $1^{\mathrm{st}}$, $2^{\mathrm{nd}}$, and $3^{\mathrm{rd}}$, and ii) blocking of outer rings such that only $0^{\mathrm{th}}$, $1^{\mathrm{st}}$ and $2^{\mathrm{nd}}$ inner rings are left in a partially blocked CADB. These types of symmetric blocking of CADB are obtained by an annular aperture, given as
\begin{equation}
H_1(r)=\left\{\begin{array}{cl}
1, & r_m \leq r \leq r_n \\
0, & \text { otherwise }
\end{array}\right. \label{eq3}
\end{equation}
Where $r=r_{n}-r_{m}$ denotes the width of annular aperture with minimum and maximum radii $r_{m}$ and $r_{n}$, respectively. The asymmetric blocking of CADB is obtained by using a sectorial aperture described as 
\begin{equation}
H_2(\phi)=\left\{\begin{array}{cl}
0, & \phi_m \leq \phi \leq \phi_n \\
1, & \text { otherwise }
\end{array}\right. \label{eq4}
\end{equation}
Where $\phi$ denotes the azimuthal angle with minimum and maximum values $\phi_m$ and $\phi_n$, respectively. 
In both cases, ``1" denotes the region with maximum transmission of light, and ``0" indicates no transmission of light. The partially blocked CADBs with symmetric and asymmetric apertures (Eqs.\,(\ref{eq3}) and (\ref{eq4})) are shown in Fig.\,\ref{fig2}. Figures\,\ref{fig2}(a)-\ref{fig2}(c) show the symmetric and asymmetric binary amplitude masks, respectively. Figures\,\ref{fig2}(d) and \ref{fig2}(e) shows the intensity distributions of partially blocked CADBs, where $0^{\mathrm{th}}$ $\&$ $1^{st}$ inner rings and the outer rings are blocked using masks given in Figs.\,\ref{fig2}(a) and \ref{fig2}(b), respectively. For the blocking of $0^{\mathrm{th}}$ $\&$ $1^{st}$ inner rings, we have used the values of $r_m=1.32$ mm and $r_n=R$, whereas for blocking of outer rings these values are used as $0$ and $1.32$ mm. Figure\,\ref{fig2}(f) shows the intensity distribution of a CADB blocked sectorially using a mask given in Fig.\,\ref{fig2}(c), where we have used the values of $\phi_m=-138^{\circ}$ and $\phi_n=138^{\circ}$.

We have numerically simulated the propagation of ideal and partially blocked CADBs under paraxial approximation in free space by using Fresnel integral as \cite{collins1970}:
\begin{eqnarray}
E(r,\phi,z)&=& \frac{k}{2\pi iz}\int_{0}^{\infty}\int_{0}^{2\pi} E(r',\phi',z) \nonumber\\
&&\times\exp\left(\frac{ik}{2z}[r^2+{r'}^{2} -2rr' \cos(\phi - \phi')]\right)\nonumber \\
&&\times r'dr' d\phi', \label{eq5}
\end{eqnarray}
where, $k=2\pi/\lambda$ with $\lambda$ is the optical wavelength, ($r',\phi'$) and ($r,\phi$) are the coordinates of initial plane ($z=0$) and observation plane ($z>0$), respectively. As analytic solution of Eq.\,(\ref{eq5}) is very difficult to find, so it is solved numerically using a fast Fourier transform method \cite{Computational_FO}.
\section{Experimental arrangement}\label{exp}
\begin{figure}[htbp]
\centering
\includegraphics[height = 3.72cm, keepaspectratio = true]{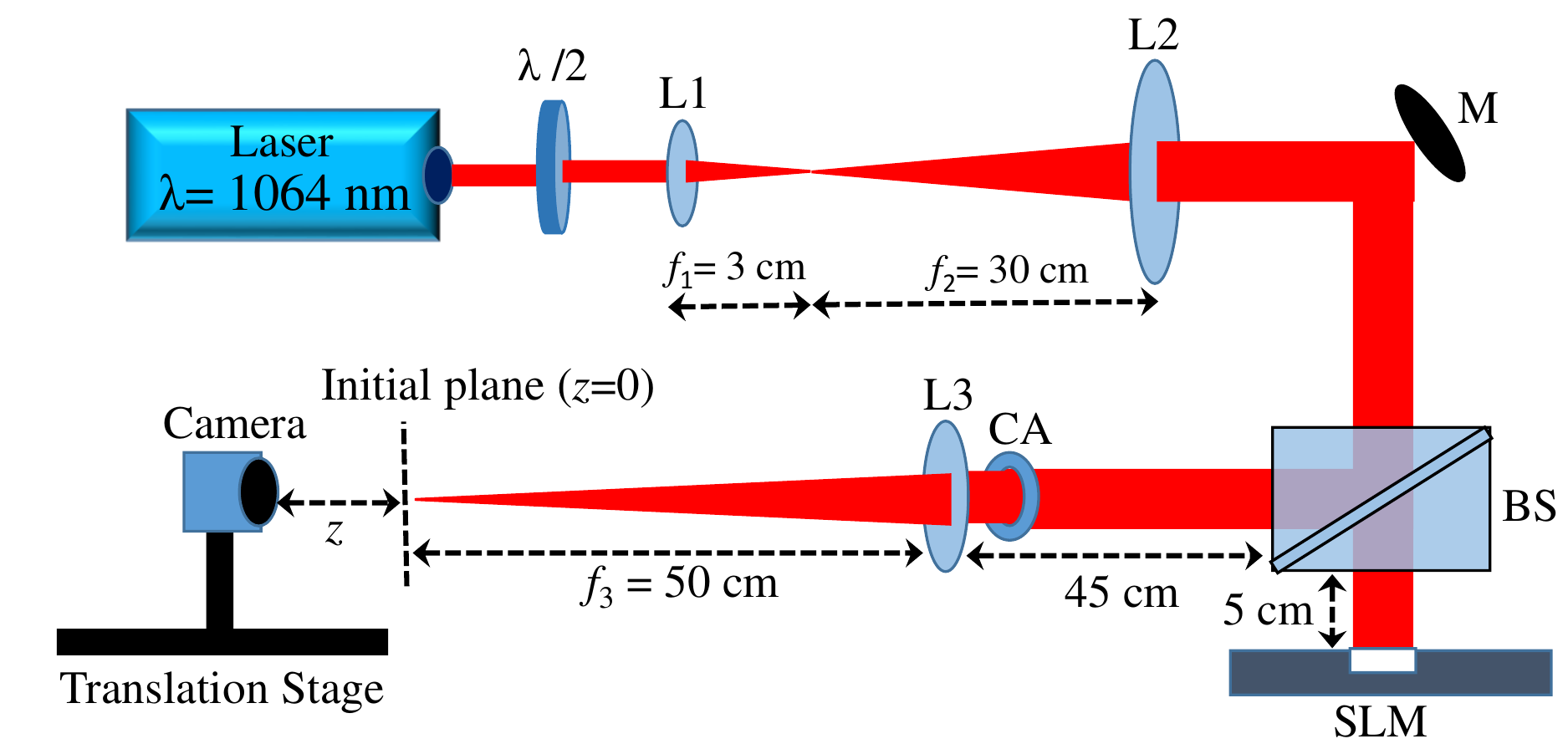}
\caption{Experimental arrangement for generation and propagation of ideal and partially blocked CADBs. $\lambda/2$: half-wave plate; $L_1$, $L_2$ and $L_3$: Plano-convex lenses of focal length $f_1=3$ cm,$f_2=30$ cm and $f_3=50$ cm, respectively; M: mirror; BS: 50:50 beam splitter; SLM: spatial light modulator; CA: circular pinhole.}
\label{fig3}
\end{figure}

The experimental arrangement for the generation and propagation of ideal CADB and partially blocked CADB is shown in Fig.\,\ref{fig3}. It consists of a diode laser, which provides an output beam with a Gaussian distribution at a wavelength of $\lambda$ = 1064 nm. The laser beam is linearly polarized with a half-wave plate, and then magnified by 10 times with a telescope made with two plano-convex lenses $L_1$~($f_1=3$ cm) and  $L_2$~($f_2=30$ cm). The magnified beam is incident perpendicularly on a phase-only spatial light modulator (SLM) using a 50:50 beam splitter (BS). The SLM consists of screen resolution $1920\times 1080$ and pixel size $8 ~\mu$m. Note, the size of an input beam is magnified such that it illuminates the whole screen of SLM. On the SLM, we impose a phase pattern that modulates amplitude and phase of an incident beam. The reflected light from SLM consists of several orders, and a suitable first order is selected by a circular pinhole (CA). The selected order is Fourier transformed with a plano-convex lens $L_{3}$, which generates a desired CADB at its focal plane (denoted as initial plane ($z=0$) in Fig.\,\ref{fig3}). The propagation of generated CADB is analyzed by using a CCD camera mounted on a translation stage.

To obtain a required phase pattern, imposing on SLM screen, for generating a desired CADB, we first Fourier transforms the electric field of a CADB (Eq.\,(\ref{eq1})) as 
\begin{equation}
E(\rho,\theta)=F[E(r,\phi,0)]=A(\rho,\theta)\exp[i\zeta(\rho,\theta)],\label{eq6}
\end{equation}
where $F$ denotes the Fourier transform operation. Now we encode the complex field $E(\rho,\theta)$ by means of a phase transmittance function (also called as phase only computer-generated hologram) to incorporate the amplitude and phase variations as
\begin{equation}
  T(x,y)=\exp[i\psi(A,\zeta)],\label{eq7}
\end{equation}
where $\psi(A,\zeta)$ takes into account the amplitude and phase variations, and corresponds to a desired phase pattern (computer generated phase hologram). To find $\psi(A,\zeta)$, we have adapted a method reported in \cite{arrizon2007}. The phase function $\psi(A,\zeta)$ with odd symmetry can be given as $\psi(A,\zeta)=f(A)\sin(\zeta)$, where an unknown function $f(A)$ needs to be determined. Hence, the transmittance function (Eq.\,(\ref{eq7})) is expressed in the Fourier series using Jacobi-Anger identity as \cite{arrizon2007}
\begin{equation}
T(x,y)=\exp[i.f(A)\sin(\zeta)]=\sum_{l=-\infty}^{\infty} J_l[f(A)]\exp(il\zeta), \label{eq8}
\end{equation}
where, $J_l$ is the first-kind of Bessel function of order $l$. Suppose that the field $E(\rho,\theta)$ is recovered from the first order term ($l=1$) in the series of Eq.\,(\ref{eq8}), provided that the following identity is fulfilled 
\begin{equation}
cA=J_{1}[f(A)],\label{eq9}    
\end{equation}
where $c$ is positive constant. The function $f(A)$ can be determined by numerical inversion of Eq.\,(\ref{eq9}). From Eq.\,(\ref{eq8}) it is clear that the transmittance contains several  orders, however, only the first-order is desired. The undesired orders ($l\neq 1$) needs to be filtered out. To do this a blazed grating is required to be added to the phase pattern, which separates different orders, and enables the selection of a desired first-order. The modified phase pattern can be written as $\psi(A,\zeta+2\pi f_x x+2\pi f_y y)$, where $f_x$ and $f_y$ are the spatial frequencies along $x-$ and $y-$directions, respectively. As a result the modulated light reflected from SLM contains several spatially separated orders, and the desired first-order is selected with a circular pinhole (CA) (Fig.\,\ref{fig3}). 

\begin{figure}
\centering
\includegraphics[height = 4.85cm, keepaspectratio = true]{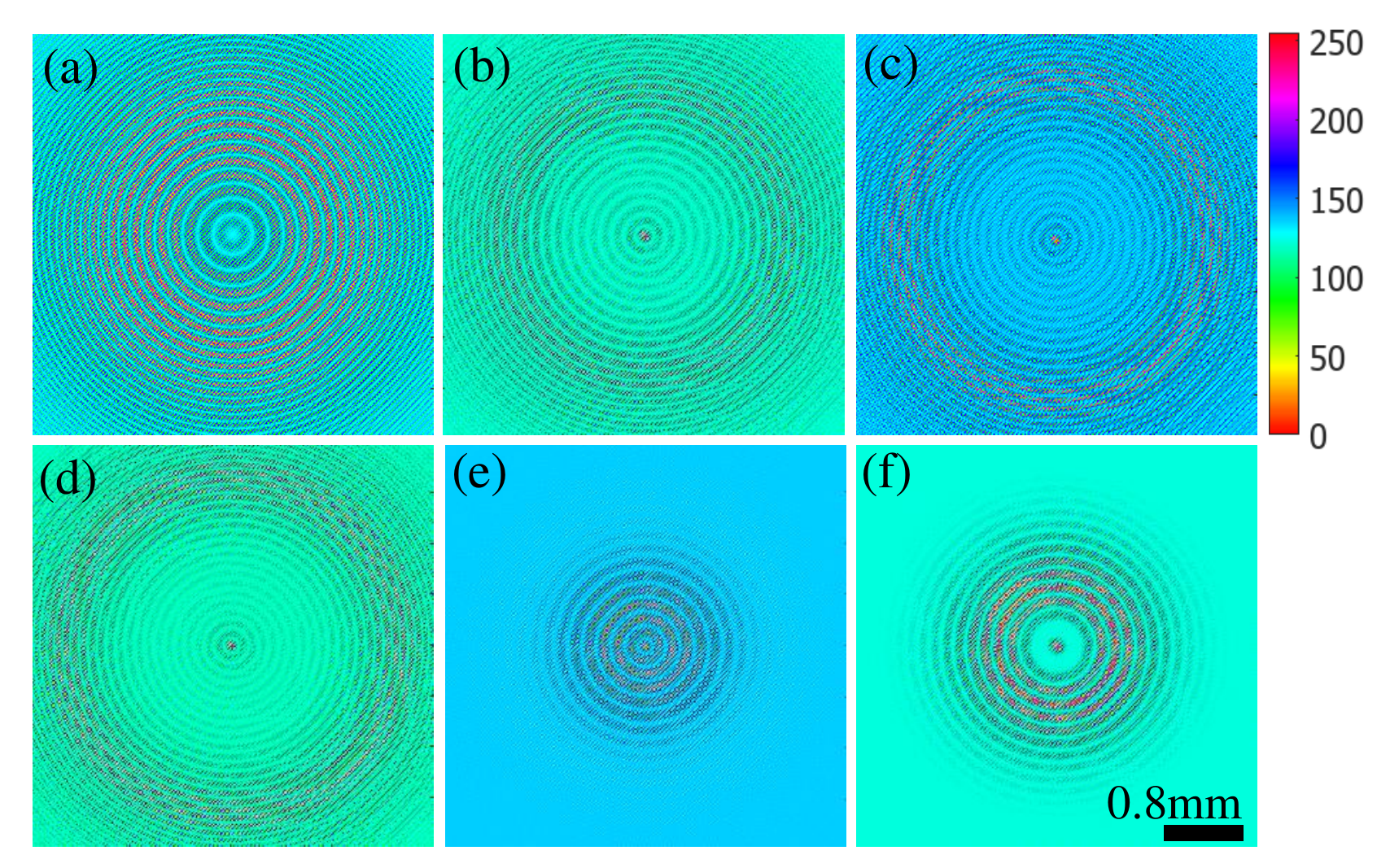}
\caption{Computer generated phase patterns for (a) ideal CADB, (b) CADB with $1^{\mathrm{st}}$ inner ring blocked, (c) CADB with two inner rings ($1^{\mathrm{st}}$ and $2^{\mathrm{nd}}$) blocked, (d) CADB with three inner rings ($1^{\mathrm{st}}$, $2^{\mathrm{nd}}$ and $3^{\mathrm{rd}}$) blocked; (e) CADB with only $0^{\mathrm{th}}$ and $1^{\mathrm{st}}$ inner rings (other outer rings are blocked), and (f) CADB with only ($0^{\mathrm{th}}$, $1^{\mathrm{st}}$ and $2^{\mathrm{nd}}$) inner rings (other outer rings are blocked). Note, the blocking of inner rings includes $0^{\mathrm{th}}$ ring.}
\label{fig4}
\end{figure}

For generating symmetrically partially blocked CADBs, the annular aperture is added inside the phase pattern (computer generated phase hologram), which enables a very precise blocking of CADB. However, for asymmetric blocking, we use a binary sector plate with $84^{\circ}$ full cone angle (Figs.\,\ref{fig2}(c) and \ref{fig2}(f)). The modified phase patterns including blazed grating as well as annular aperture for generating ideal CADB and partially blocked CADB are shown in Fig.\,\ref{fig4}.

Figure\,\ref{fig4}(a) shows the phase pattern for generating ideal CADB. Figures\,\ref{fig4}(b)-\ref{fig4}(c) show the phase patterns for generating CADB with symmetrically blocked $1^{\mathrm{st}}$ inner ring, two inner rings ($1^{\mathrm{st}}$ and $2^{\mathrm{nd}}$), and three inner rings ($1^{\mathrm{st}}$, $2^{\mathrm{nd}}$ and $3^{\mathrm{rd}}$), respectively. Note, for inner blocking $0^{\mathrm{th}}$ ring is included. Figures\,\ref{fig4}(e) and \ref{fig4}(f) show the phase patterns for CADB with symmetrically blocked outer rings and having only $1^{\mathrm{st}}$ inner ring and two inner rings ($1^{\mathrm{st}}$ and $2^{\mathrm{nd}}$) (includes $0^{\mathrm{th}}$ inner ring in both cases), respectively. 
\section{Results and discussions}\label{results}
\subsection{Symmetrically blocked CADBs}
\begin{figure}[htbp]
\centering
\includegraphics[height = 7.5cm, keepaspectratio = true]{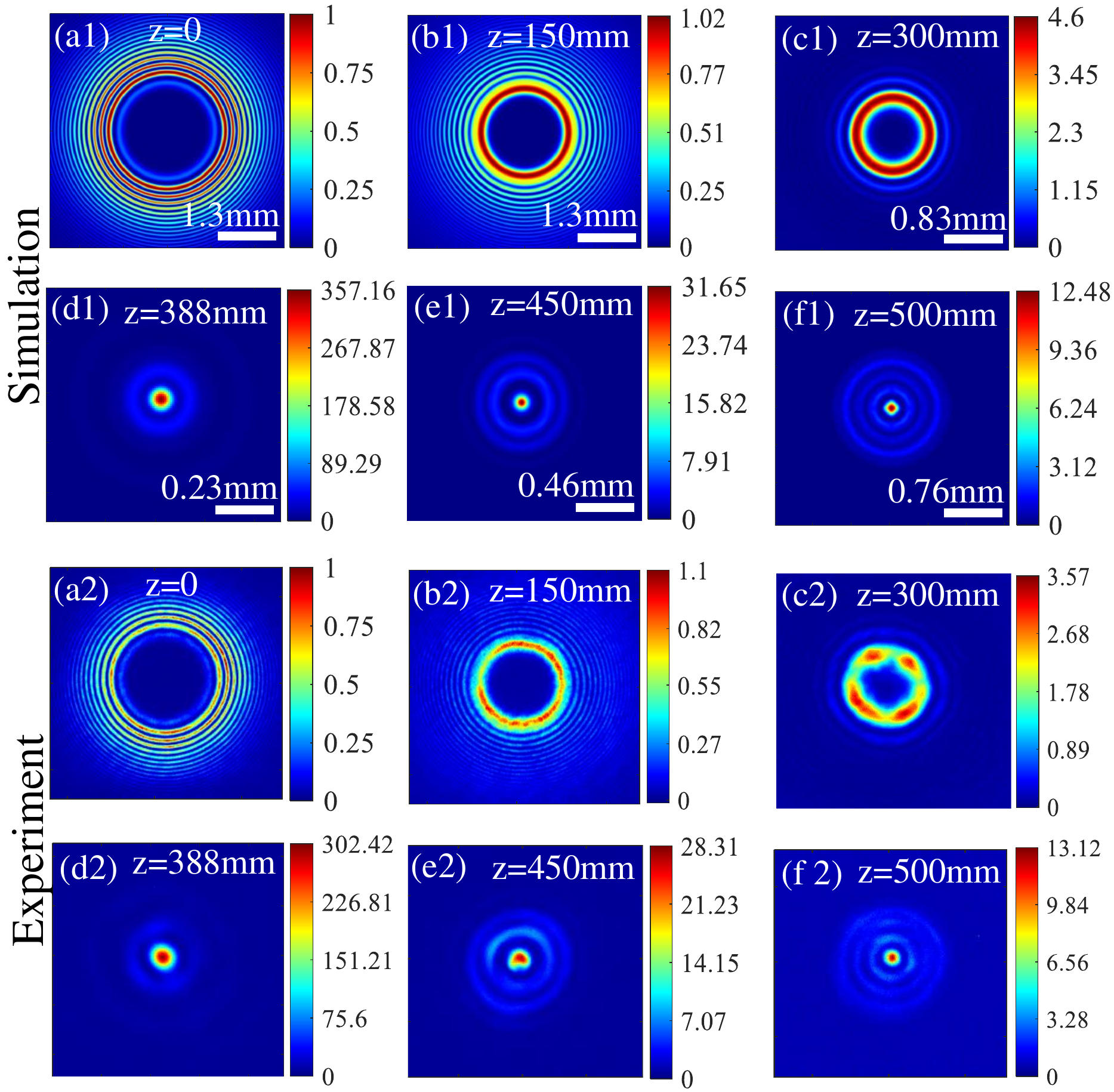}
\caption{Simulated and experimental intensity distributions of an ideal CADB at different propagation distance (a1, a2) $z=0$, (b1, b2) $z=150$ mm, (c1, c2) $z=300$ mm, (d1, d2) $z=388$ mm, (e1, e2) $z=450$ mm, and (f1, f2) $z=500$ mm. The parameter values are taken as $r_0=1$ mm, $w_0=0.1$ mm, $a=0.1$ and $\lambda=1064$ nm. The maximum value in colorbar represents the k-value. Note, the scalebars in experimental results are the same at respective $z$.}
\label{fig5}
\end{figure}

First we have investigated the autofocusing and self-healing of symmetrically blocked CADBs. The results are compared with an ideal CADB to analyze the effect of blocking. The simulated results of propagation of CADB are obtained by solving numerically Eq.\,(\ref{eq5}). The simulated and experimental results of an ideal CADB with parameter values of $r_0=1$ mm, $w_0=0.1$ mm, $\lambda=1064$ nm and $a=0.1$ are shown in Fig.\,\ref{fig5}. 
As evident, the ideal CADB consists of multiple rings (Figs.\,\ref{fig5}(a1) and \ref{fig5}(a2)), and as the beam propagates the intensity from outer rings moves towards the inner rings, and finally at a distance of $z=388$ mm most of the intensity becomes tightly focused in the form of a single peak with maximum intensity at the center. This distance at which single peak with maximum intensity is obtained, called the autofocusing distance $z_{af}$ (Eq.\,(\ref{eq2})). Note, in these figures a maximum value in colorbar represents the k-value. The k-value increases with $z$, and becomes maximum at $z_{af}$, and after that it decreases. The simulated and experimental k-value at $z_{af}$ are found to be 357.16 (Fig.\,\ref{fig5}(d1)) and 302.42 (Fig.\,\ref{fig5}(d2)), as predicted earlier \cite{zang2022}. A small discrepancy between the simulated and experimental k-value is observed, which is anticipated by the non-similarities in outer rings as well as non-uniform intensity distribution azimuthally on the rings. Particularly, in the experiment the tail part (weak intensity regions) of an input Gaussian beam on the SLM is not modulated very well, and due to that outer low intensities rings in the generated CADB are affected. In calculation of k-value, outer rings play an important role. Further, the experimentally generated CADB does not consist of uniform intensity azimuthally on the rings due to imperfections in system. The understanding of autofocusing in CADB has been provided by an analogy to the Fresnel zone plate \cite{zang2022}. After the autofocusing, intensity again flows outwards in the form of generating rings, but a bright spot remains at the center with a reduced k-value. The simulation and experiment show a reasonably good agreement.    
\subsubsection{Blocking inner rings of CADB} \label{inblock}
For a partial blocking from inside, we have blocked CADB from the center ($r=0$) up to $1^{\mathrm{st}}$, $2^{\mathrm{nd}}$ and $3^{\mathrm{rd}}$ inner rings. The dark (zero intensity) central region of partially blocked CADB increases with the number of blocked inner rings. The amount of blocking is quantified by the method of diffraction efficiency ($\eta$), which is defined as the ratio of intensity in the partially blocked beam to the total intensity of the ideal beam. For the CADB blocked up to $0^\mathrm{th}$, $1^{\mathrm{st}}$, $2^{\mathrm{nd}}$ and $3^{\mathrm{rd}}$ inner rings, the diffraction efficiencies are found to be $96.29\%$, $81.57\%$, $70.13\%$ and $60.72\%$, respectively. As blocking of $1^{\mathrm{st}}$, $2^{\mathrm{nd}}$ and $3^{\mathrm{rd}}$ inner rings removes a significant amount of intensity, so it is interesting to check the propagation properties corresponding to each case with the increase of dark central region. Note, $0^{\mathrm{th}}$ inner ring consists of a very low intensity, so it is always blocked while blocking the first ($1^{\mathrm{st}}$), two ($1^{\mathrm{st}}$ and $2^{\mathrm{nd}}$ ), three ($1^{\mathrm{st}}$, $2^{\mathrm{nd}}$ and $3^{\mathrm{rd}}$) inner rings. 

\begin{figure}[htbp]
\centering
\includegraphics[height = 7.5cm, keepaspectratio = true]{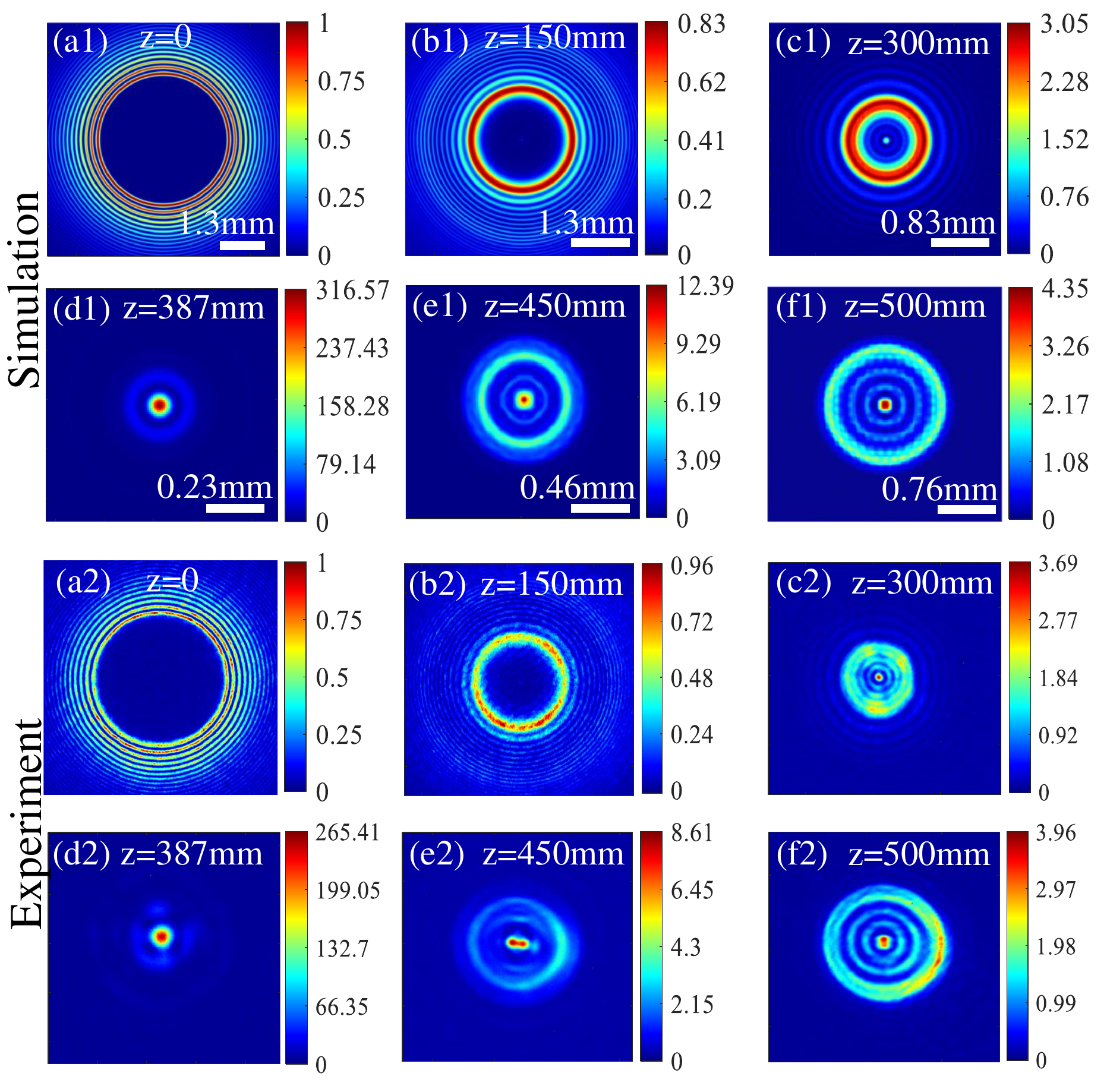}
\caption{Simulated and experimental intensity distributions of CADB with $1^{\mathrm{st}}$ inner ring blocked, at different propagation distance (a1, a2) $z = 0$, (b1, b2) $z=150$ mm, (c1, c2) $z=300$ mm, (d1, d2) $z=387$ mm, (e1, e2) $z=450$ mm, and (f1, f2) $z=500$ mm. The maximum value in colorbar represents the k-value. Note, the scalebars in experimental results are the same at respective $z$.}
\label{fig6}
\end{figure}

First, we have blocked $1^{\mathrm{st}}$ inner ring (Fig.\,\ref{fig1}(b)) of CADB at initial plane $z=0$ using an annular aperture (Eq.\,\ref{eq3}, Figs.\,\ref{fig2}(a) and \ref{fig2}(d)) with values of $r_{m}=1.32 $mm and $r_{n}=R$, and propagates it for distances $z>0$. The simulated and experimental results are shown in Fig.\,\ref{fig6}. 
Figures\,\ref{fig6}(a1) and \ref{fig6}(a2) show the simulated and experimental intensity distributions of partially blocked CADB, indicating that the $1^{\mathrm{st}}$ inner ring is blocked. The diffraction efficiency is reduced to 81.57\%, hence, 18.43\% of total intensity is blocked. Figures\,\ref{fig6}(b1)-\ref{fig6}(f1) (simulation) and Figs.\,\ref{fig6}(b2)-\ref{fig6}(f2) (experiment) show the intensity distributions of partially blocked CADB at propagation distances $z=150$ mm, $300$ mm, $387$ mm, $450$ mm, and $500$ mm, respectively. It is evident that as the partially blocked CADB propagates the intensity from outer rings flows towards the inner rings (shown with increased intensity), and it autofocuses at a distance of $z_{af}=387$ mm, after that it again defocuses and intensity redistributes in multiple rings, while retaining a bright central spot. Similar to an ideal CADB (Fig.\,\ref{fig5}), partially blocked CADB shows the same propagation behavior, and autofocuses approximately at the same distance $z_{af}=387$ mm. Further, the k-value (shown in colorbar) of partially blocked CADB also shows the same trend as observed for an ideal CADB. At $z_{af}$ the simulated and experimental k-value are found to be $\sim316.6$ and $\sim265.4$, respectively, which are smaller than that of an ideal CADB (Figs\,\ref{fig5}(d1) and \ref{fig5}(d2)). This indicates that the k-value depends on the amount of blocking of CADB, as the blocking of first inner ring leads to a reduced diffraction efficiency of $\eta=81.57\%$. The discrepancy between the experimental and simulated k-value is due to the same reasons as explained above in the case of an ideal CADB. Further, like an ideal CADB, the similar observations of partially blocked CADB indicate that the blocked inner ring self-heals by redistribution of intensity from other nearby rings, accordingly the beam retains similar propagation properties. 
%In particular, after the autofocusing the rings in ideal and blocked CADB appears approximately at the same locations, however, the intensity is found to be different in the later case.

\begin{figure*}[htbp]
\centering
\includegraphics[height = 8cm, keepaspectratio = true]{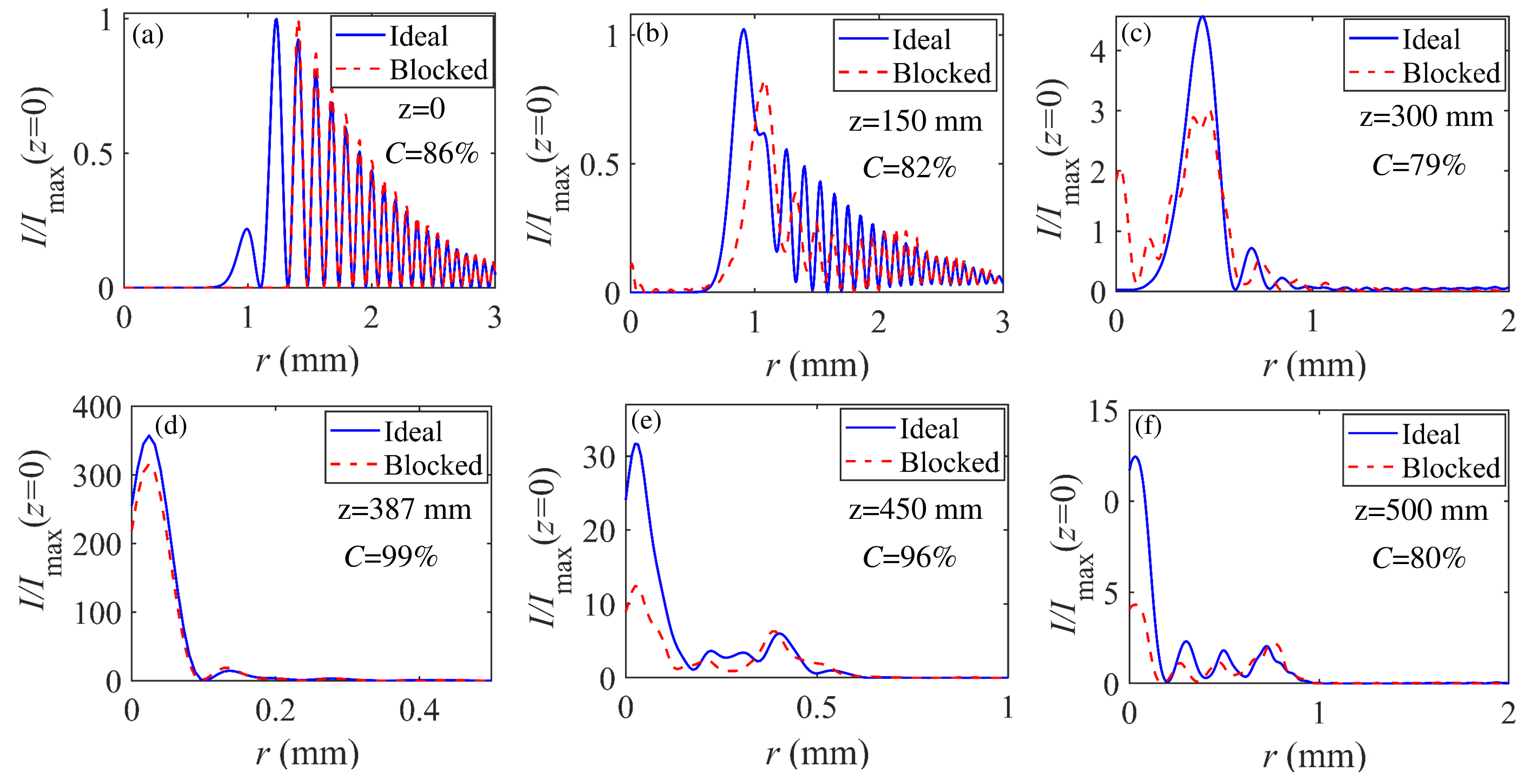}
\caption{Normalized intensity cross sections of ideal CADB (blue solid curve) and blocked CADB ($1^{\mathrm{st}}$ inner ring blocked) (red dashed curve) at various propagation distances (a) $z=0$, (b) $z=150$ mm, (c) $z=300$ mm, (d) $z=387$ mm, (e) $z=450$ mm, and (f) $z=500$ mm. The cross-sections are taken along the horizontal axis in Figs.\,\ref{fig5}(a1)-\ref{fig5}(f1) and Figs.\,\ref{fig6}(a1)-\ref{fig6}(f1). $C$ denotes the value of overlap integral (Eq.\,(\ref{eq10})).}
\label{fig7}
\end{figure*}

In order to gain a better understanding of self-healing, we have compared the normalized intensity cross-sections of partially blocked CADB (red dashed curve) and ideal CADB (blue solid curve) at various propagation distances, as shown in Figs.\,\ref{fig7}(a)-\ref{fig7}(f). As evident, during the propagation the maximum intensity ring, which is blocked at $z=0$, emerges at $z=150$ mm by flowing the intensity from other neighboring rings (red dashed curve in Fig.\,\ref{fig7}(b)) at a slightly different spatial position, and as the beam propagates further, it becomes more similar to an ideal CADB. In particular, at $z_{af}$, both ideal and partially blocked CADB show high similarities, but with a reduced peak intensity in the blocked case (Fig.\,\ref{fig7}(d)). After $z_{af}$, the intensity again redistributes, and rings are formed in both cases. The number of rings in both cases are found to be the same at $z=500$ mm, as shown in Fig.\,\ref{fig7}(f), however, the intensity on the rings differs. The results clearly show that the partially blocked CADB self-heals, and it becomes maximum at $z_{af}$.

To quantify the self-healing, we have calculated an overlap integral that measures the similarities between the ideal and self-healed beams \cite{dev2021divergence}.
\begin{equation}
C(z)=\frac{\int \int I_o(x,y,z)I_s(x,y,z)dxdy}{\sqrt{\int \int I_o^{2}(x,y,z)dxdy}\sqrt{\int \int I_s^{2}(x,y,z)dxdy}}, \label{eq10}
\end{equation}
where $I_o(x,y,z)$ and $I_s(x,y,z)$ denote the intensities of ideal and self-healed beams, respectively. We have calculated the value of $C$ at different propagation distances for partially blocked CADB, shown in Fig.\,\ref{fig7}. The values are found to be $C=86 \%$, $82\%$, $79\%$, $99\%$, $96\%$, and $80\%$ at the propagation distances $z=0$, $150$ mm, $300$ mm, $387$ mm, $450$ mm, and $500$ mm, respectively. Due to partial blocking of CADB at $z=0$, the value of $C$ at $z=0$ is found to be reduced. Initially the value of $C$ fluctuates due to redistribution of intensity for self-healing of blocked parts of the beam, and it becomes maximum at $z_{af}$, and after that it again fluctuates and slowly attains a fixed value of $\sim 90\%$ for longer propagation distances $z>500$ (shown later).

\begin{figure}[htbp]
\centering
\includegraphics[height = 7.8cm, keepaspectratio = true]{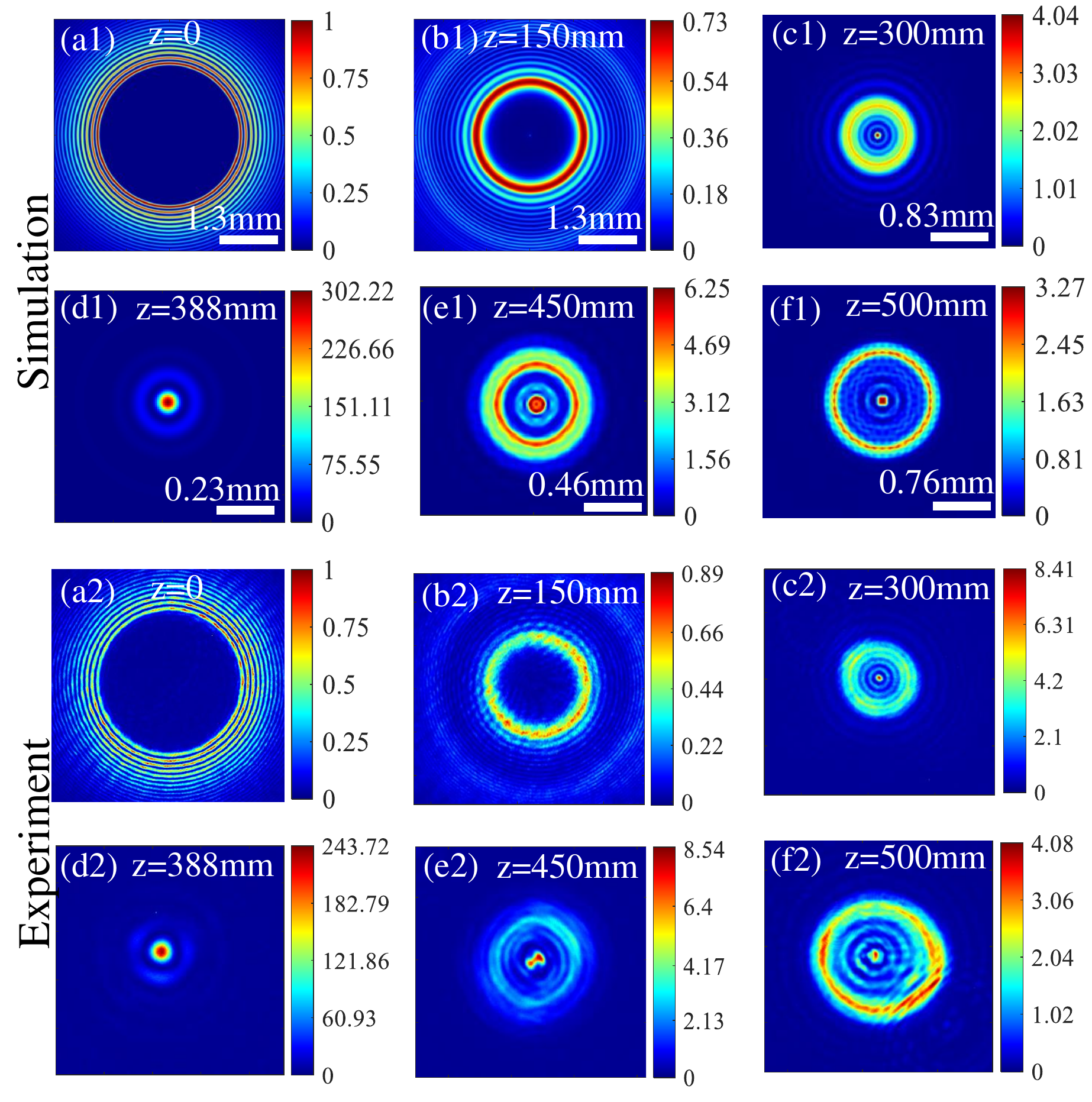}
\caption{Simulated and experimental intensity distributions of CADB with two inner rings ($1^{\mathrm{st}}$ and $2^{\mathrm{nd}}$) blocked, at different propagation distances (a1, a2) $z = 0$, (b1, b2) $z=150$ mm, (c1, c2) $z=300$ mm, (d1, d2) $z=388$ mm, (e1, e2) $z=450$ mm, and (f1, f2) $z=500$ mm. The maximum value colorbar represents the k-value. The parameter values are kept the same as in Fig.\,\ref{fig5}. Note, the scalebar in experimental results are the same at respective $z$.}
\label{fig8}
\end{figure}

Further, we have blocked two inner rings ($1^{\mathrm{st}}$ and $2^{\mathrm{nd}}$) of CADB using an annular aperture (Eq.\,(\ref{eq3})) with values of $r_m=1.48$ mm and $r_n=R$. This leads to a reduction of diffraction efficiency $\eta=70.13\%$, and $29.87\%$ of total intensity is blocked. The simulated and experimental results of propagation of partially blocked CADB are shown in Fig.\,\ref{fig8}. Figures\,\ref{fig8}(a1) and \ref{fig8}(a2) show the simulated and experimental intensity distributions of CADB with two inner rings blocked, respectively. The intensity distributions of partially blocked CADB at different propagation distances are shown in Figs.\,\ref{fig8}(b1)-\ref{fig8}(f1) (simulation) and Figs.\,\ref{fig8}(b2)-\ref{fig8}(f2) (experiment). It is evident that the CADB blocked with two inner rings again autofocuses at a distance of $z_{af}=388$ mm (Figs.\,\ref{fig8}(d1) and \ref{fig8}(d2)), showing similar observation as obtained for an ideal CADB (Figs.\,\ref{fig5}(d1) and \ref{fig5}(d2)) and CADB blocked with first inner ring (Figs.\,\ref{fig8}(d1) and \ref{fig8}(d2)). Further, like an ideal CADB and CADB blocked with first inner ring, a similar trend in the k-value (shown in colorbar) is again observed, where k-value increases with the distance $z$, and reaches maximum value at $z_{af}$, and after that it again decreases. At $z_{af}$ the maximum k-value is found to be $\sim302$ in the simulation (Fig.\,\ref{fig8}(d1)) and $\sim244$ in the experiment (Fig.\,\ref{fig8}(d2)). The maximum k-value is found to be reduced as compared to CADB blocked with first inner ring, indicating that the k-value depends on the amount of blocking. After $z_{af}$, the redistribution of intensity on rings is also found to be different than that of an ideal CADB and CADB blocked with first inner ring. The values of overlap integral are found to be $C=65\%$, $68\%$, $74\%$, $98\%$, $92\%$, and $77\%$ at the propagation distances  $z=0$, $150$ mm, $300$ mm, $388$ mm, $450$ mm, and $500$ mm, respectively, which shows a similar trend as obtained for CADB blocked with first inner ring. It shows that partially blocked CADB self-heals, and maximum self-healing again occurs at $z_{af}$.

Further, we have blocked three inner rings ($1^{\mathrm{st}}$, $2^{\mathrm{nd}}$, and $3^{\mathrm{rd}}$) in CADB using an annular aperture (Eq.\,(\ref{eq3})) with values of $r_m=1.616$ mm and $\ r_n=R$, and checked the autofocusing and self-healing abilities. Blocking of three inner rings leads to a reduction of diffraction efficiency $\eta=60.72\%$.
\begin{figure}[htbp]
\centering
\includegraphics[height = 7.5cm, keepaspectratio = true]{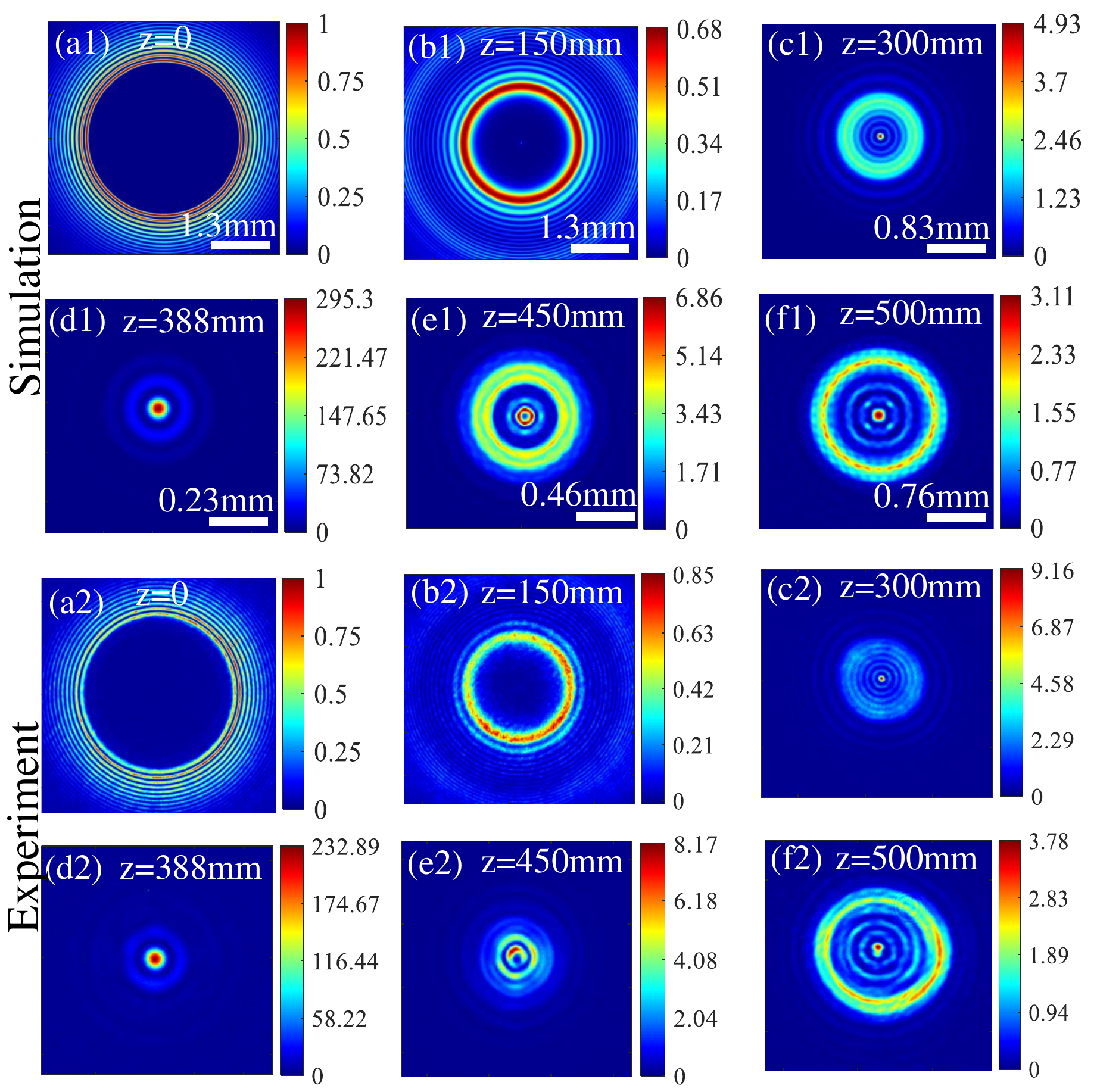}
\caption{Simulated and experimental intensity distributions of CADB blocked with three inner rings ($1^{\mathrm{st}}$, $2^{\mathrm{nd}}$ and $3^{\mathrm{rd}}$) at different propagation distances (a1, a2) $z = 0$, (b1, b2) $z=150$ mm, (c1, c2) $z=300$ mm, (d1, d2) $z=388$ mm, (e1, e2) $z=450$ mm, and (f1, f2) $z=500$ mm. The maximum value on colorbar represents the k-value. The parameter values are kept the same as in Fig.\,\ref{fig5}. Note, the scalebars in experimental results are the same at respective $z$.}
\label{fig9}
\end{figure}
Figures\,\ref{fig9}(a1) and \ref{fig9}(a2) show the simulated and experimental intensity distributions of CADB blocked with three inner rings at $z=0$, respectively, indicating a large dark central region. The simulated and experimental intensity distributions of partially blocked CADB at different propagation distances are shown in Figs.\,\ref{fig9}(b1)-\ref{fig9}(f1) and Figs.\,\ref{fig9}(b2)-\ref{fig9}(f2), respectively. As the partially blocked CADB propagates, the autofocusing is again observed at the same distance $z_{af}=388$ mm (Fig.\,\ref{fig9}(d1) and \ref{fig9}(d2)). This indicates that even though a large part of the CADB is blocked, the beam does not loose its autofocusing abilities, and $z_{af}$ remains unchanged. After the autofocusing the intensity again redistributes and rings are formed, however, intensity on the rings appears differently from the previous cases (Figs.\,\ref{fig5}, \ref{fig6}, and \ref{fig8}). The k-value again increases initially with the propagation distance (shown in colorbar) and reaches a maximum value of $\sim 295$ in the simulation and $\sim 233$ experimentally at $z_{af}$, and after that it again decreases. The discrepancy between the simulated and experimental k-value again appears due to the same reason as explained earlier. Due to large amount of blocking the maximum k-value at $z_{af}$ is found to be smallest as compared to an ideal CADB, and CADB blocked with first inner ring and two inner rings. To check the self-healing effects, we have calculated the overlap at different propagation distances, and the values are found to be $C=49\%$, $57\%$, $67\%$, $93\%$, $88\%$, and $76\%$ at the propagation distances  $z=0$, $150$ mm, $300$ mm, $388$ mm, $450$ mm, and $500$ mm respectively. The variation of $C$ with distance shows a similar trend as observed earlier, and the maximum value of $C$ occurs at $z_{af}$, indicating a maximum self-healing of partially blocked CADB.

\begin{figure}[htbp]
\centering
\includegraphics[height = 6.2 cm, keepaspectratio = true]{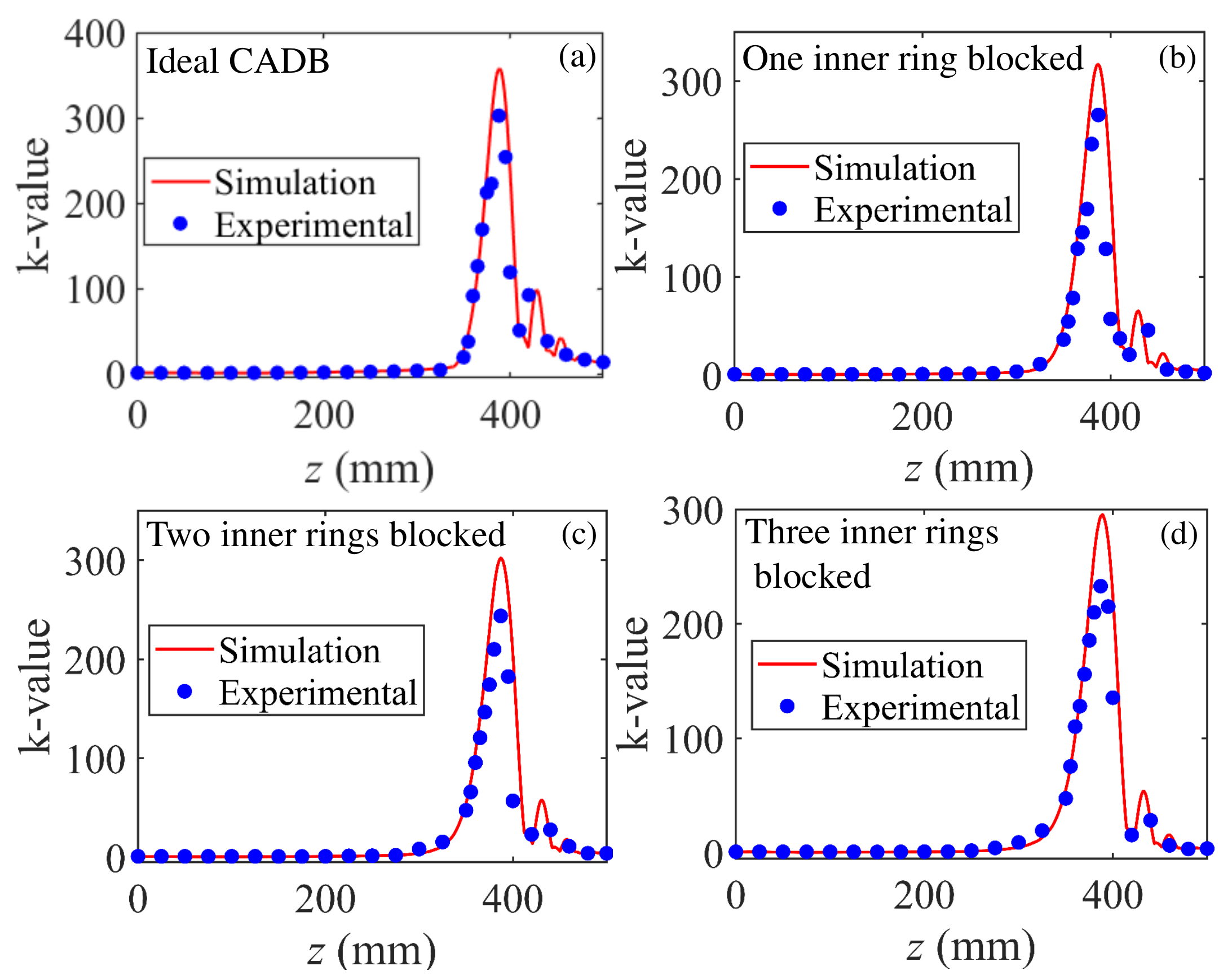}
\caption{Experimental (blue filled circles) and simulated (red solid curve) k-value as a function of propagation distance $z$ for (a) ideal CADB, (b) CADB with first inner ring blocked, (c) CADB with two inner rings blocked, and (d) CADB with three inner rings blocked.}
\label{fig10}
\end{figure}

A detailed comparison of simulated and experimental k-value as a function of propagation distance for an ideal CADB and CADB blocked with first inner ring, two inner rings, and three inner rings is shown in Fig.\,\ref{fig10}. The simulation and experimental results are shown in red solid curve and blue solid circles, respectively. As evident, the k-value increases with the propagation distance and reaches a maximum and after that it decreases with some small oscillations in its value. In all the cases a maximum k-value occurs at  $z_{af}\simeq388$ mm. In particular, the k-value is found to be largest in an ideal CADB (Fig.\,\ref{fig10}(a)), and it decreases as the number of blocked inner rings increases (Fig.\,\ref{fig10}(b)-\ref{fig10}(d)). Autofocusing property of CADB is determined by an increase in the k-value with $z$, and it's strength depends on the maximum k-value. It is evident that an ideal CADB possesses strong autofocusing ability, and it decreases with the increase in the number of blocked inner rings. Further, the abruptness of autofocusing depends on how fast the k-value varies (increase and decrease) just before and after the $z_{af}$. In all these cases, the k-value sharply increase and decrease around the $z_{af}$, indicating that ideal CADB and CADB blocked with inner rings possess an abrupt autofocusing.

Unlike CADB, in a common CAB the maximum k-value increases by increasing number of blocked inner rings \cite{li2014}. In particular, the k-value is found to be enhanced by a factor of 2.7, 3.38, and 3.79, for blocking of first, two and three inner rings, respectively. However, for CADB, the k-value is reduced by a factor of 0.88, 0.84, and 0.82 for blocking of first, two, and three inner rings respectively.

\begin{figure}[htbp]
\centering
\includegraphics[height = 5cm, keepaspectratio = true]{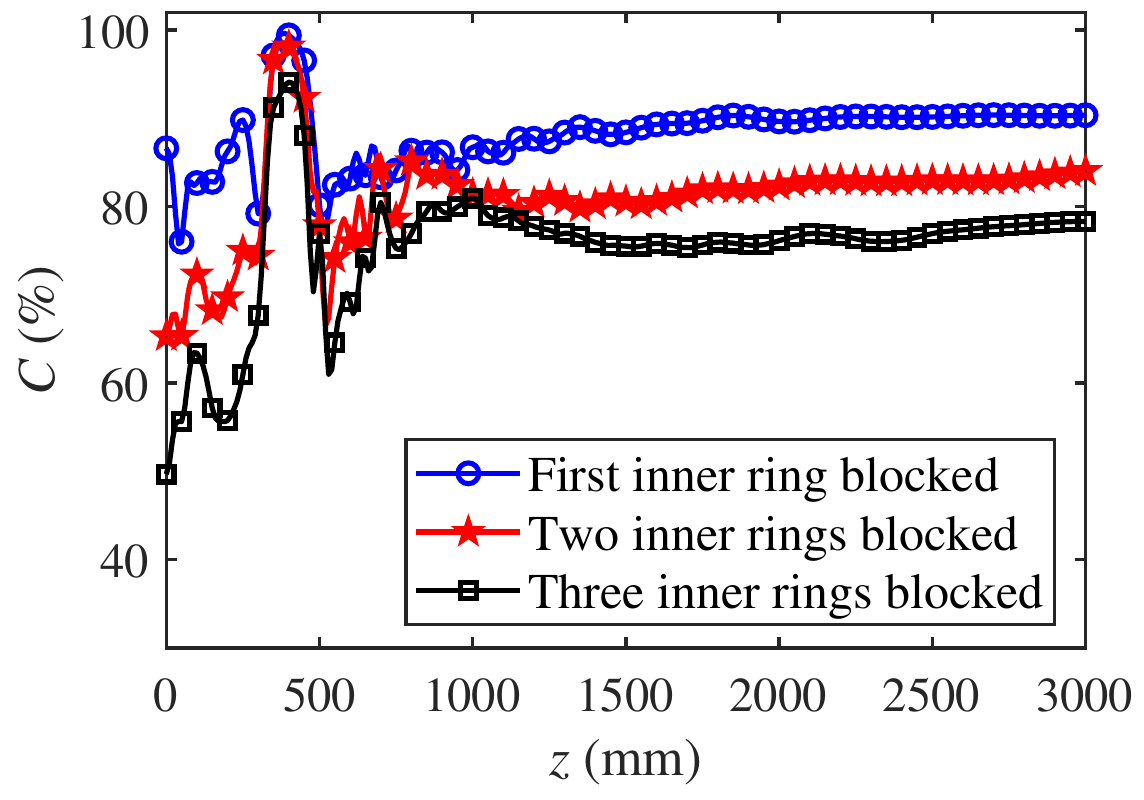}
\caption{The overlap integral ($C$) as a function of propagation distance ($z$) for CADB blocked with first inner ring (blue solid curve with circles), two inner rings (red solid curve with stars) and three inner rings (black solid curve with squares).} 
\label{fig11}
\end{figure}

Further, a detailed quantification of self-healing of partially blocked CADB is performed by calculating an overlap integral (Eq.\,\ref{eq10}) as a function of propagation distance, as shown in Fig.\,\ref{fig11}. The blue solid curve with circles, red solid curve with stars, and black solid curve with squares denote the results for CADB blocked with first inner ring, two inner rings, and three inner rings, respectively.
At $z=0$ due to blocking of first, two and three inner rings, the reduced values of overlap $C$ are found to be $\sim 86\%$, $\sim 65\%$ and $\sim 49\%$, respectively. As the partially blocked CADB propagates for $z>0$, initially the value of $C$ fluctuates due to redistribution of intensity, and then increases to a maximum value at $z_{af}$, and after that it again decreases and fluctuates, and finally it stabilizes to a fixed value. Due to the redistribution of intensity, blocked parts of CADB self-heals, which gives rise to an increase in the value $C$. At $z_{af}$, the maximum values of $C$ are found to be $\sim99\%$, $\sim98\%$ and $\sim93\%$, for CADB blocked with first inner ring, two inner rings, and three inner rings, respectively, indicating the maximum self-healing of partially blocked CADBs. It clearly shows that the self-healing distance of CADB blocked with inner rings remains invariant with respect to the amount of blocking. However, the value $C$ is found be largest for small blocking. After the autofocusing, at a large distance $z>1000$ mm, the values of $C$ are found to be $\sim90\%$, $\sim84\%$ and $\sim78\%$ corresponding to CADB blocked with first, two and three inner rings, respectively. It clearly shows that the CADB exhibits good self-healing abilities. 

\subsubsection{Blocking outer rings in CADB}
\begin{figure}[htbp]
\centering
\includegraphics[height = 8.0cm, keepaspectratio = true]{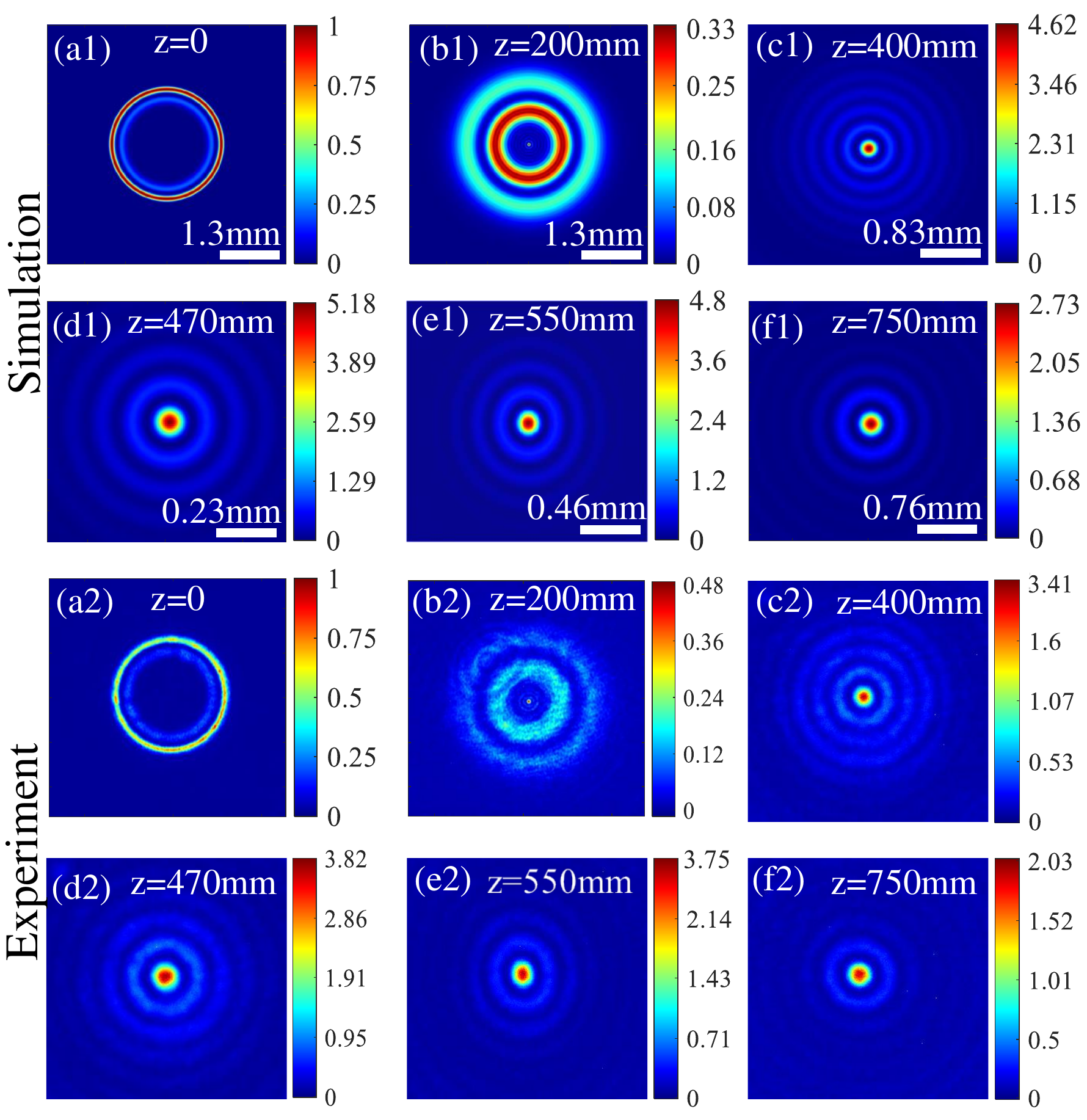}
\caption{Simulated and experimental intensity distributions of CADB with outer rings blocked (has only $0^{\mathrm{th}}$ and $1^{\mathrm{st}}$ inner rings), at different propagation distances (a1, a2) $z=0$, (b1, b2) $z=200$ mm, (c1, c2) $z=400$ mm, (d1, d2) $z=470$ mm, (e1, e2) $z=550$ mm, and (f1, f2) $z=750$ mm. The maximum value in colorbar represents the k-value. The parameter values are taken the same as in Fig.\ref{fig5}. Note, the scalebars in experimental results are the same at respective $z$.}
\label{fig12}
\end{figure}

Next, we have investigated the autofocusing and self-healing of CADB with outer rings blocked at $z=0$. To do that we use a binary annular aperture (Fig.\,\ref{fig2}(b)) that blocks all the outer rings, and leaves only the $0^{\mathrm{th}}$ and $1^{\mathrm{st}}$ inner rings in CADB (Fig.\,\ref{fig2}(e)). The partially blocked CADB is propagated for $z>0$. The simulated and experimental results are shown in Fig.\,\ref{fig12}. Figures\,\ref{fig12}(a1) and \ref{fig12}(a2) show the simulated and experimental intensity distributions of partially blocked CADB at $z=0$, respectively, indicating that it has only $0^{\mathrm{th}}$ and $1^{\mathrm{st}}$ inner rings. This leads to a reduced diffraction efficiency of $\eta=18.43\%$, indicating that a large amount of intensity ($81.57\%$) has been blocked. Figures\,\ref{fig12}(b1)-\ref{fig12}(f1) and Figs.\,\ref{fig12}(b2)-\ref{fig12}(f2) show the simulated and experimental intensity distributions of partially blocked CADB at different propagation distances $z=200$ mm, $400$ mm, $470$ mm, $550$ mm, and $750$ mm, respectively. As evident, partially blocked CADB shows the autofocusing at a distance of $z_{af}=470$ mm. The autofocusing distance is found to be longer than that of an ideal CADB (Fig.\,\ref{fig5}(d1) and \ref{fig5}(d2)) and CADB blocked with inner rings (Figs.\,\ref{fig6}(d1), \ref{fig8}(d1) and \ref{fig9}(d1)). Further, the k-value (shown in colorbar) first increases with the distance and reaches a maximum value at $z_{af}$ and after that it again decreases. The maximum k-value at $z_{af}$ is found to be $5.18$ in the simulation (Fig.\,\ref{fig12}(d1)) and $3.82$ in the experiment (Fig.\,\ref{fig12}(d2)), respectively. These maximum values are found to be much smaller than that of an ideal CADB (Fig.\,\ref{fig5}) and CADB blocked with inner rings (Figs.\,\ref{fig6}, \ref{fig8} and \ref{fig9}). For checking the self-healing, we have calculated the overlap values, which are found to be $72\%$, $75\%$, $91\%$, $86\%$, $78\%$ and $62\%$ corresponding to $z=0$, $200$ mm, $400$ mm, $470$ mm, $550$ mm, and $750$ mm. It shows that the partially blocked CADB again self-heals, and it becomes maximum at $z_{af}$. Unlike the previous cases of an ideal CADB and CADB blocked with inner rings, after $z_{af}$ the ring structure of self-healed beam remains almost the same (Figs.\,\ref{fig12}(e1)- \ref{fig12}(f1) and Figs.\,\ref{fig12}(e2)-\ref{fig12}(f2)). It clearly depicts that outer rings play an important role in the autofocusing and self-healing properties of CADB.

\begin{figure*}[htbp]
\centering
\includegraphics[height = 8cm, keepaspectratio = true]{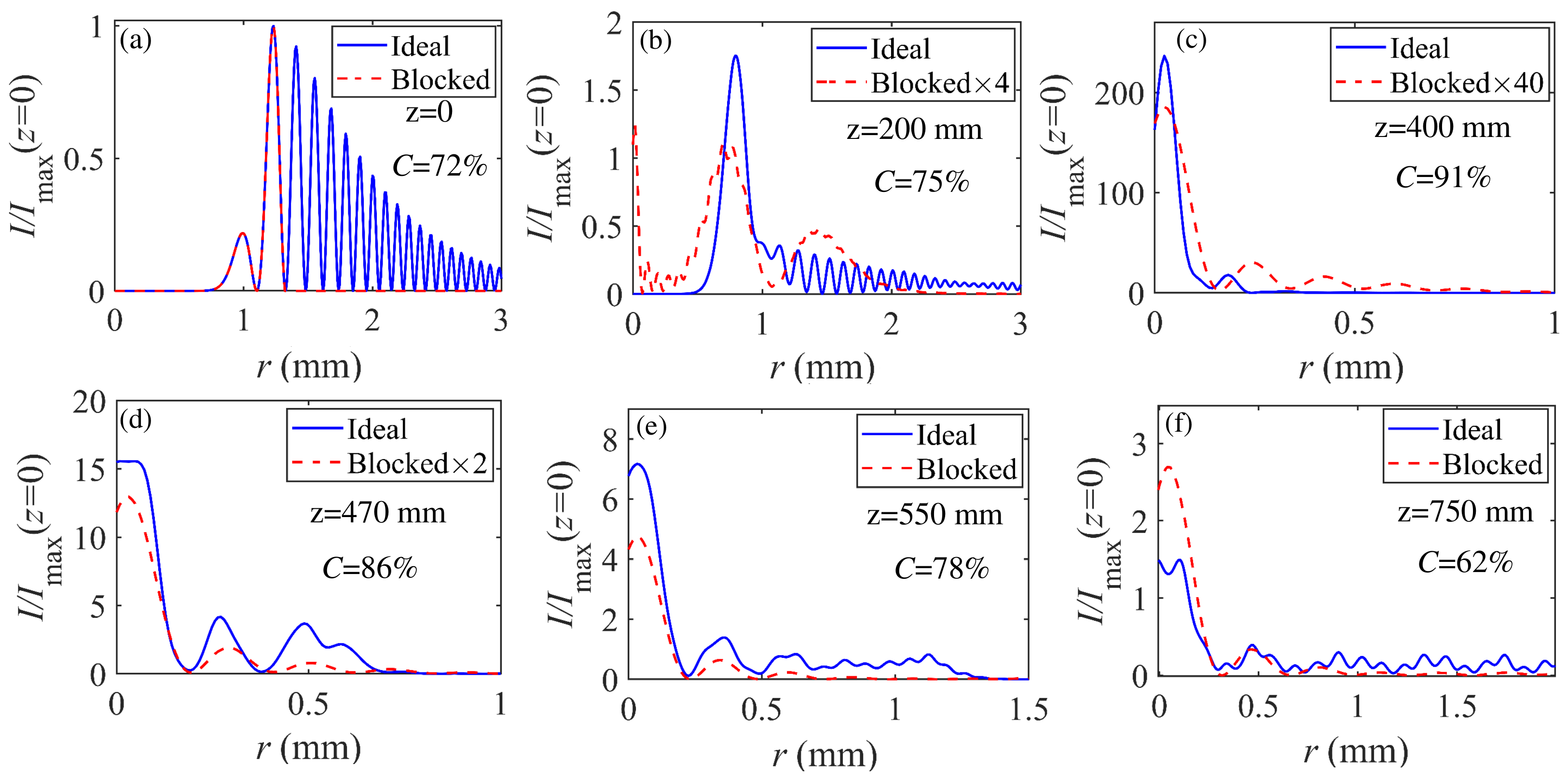}
\caption{(a) Simulated intensity cross-sections of ideal CADB (blue solid curve) and blocked CADB (contains only $0^{\mathrm{th}}$ and $1^{\mathrm{st}}$ inner rings) (red dashed curve) at various propagation distances (a) $z=0$, (b) $z=200$ mm, (c) $z=400$ mm, (d) $z=470$ mm, (e) $z=550$ mm, and (f) $z=750$ mm. The cross-sections are taken along the horizontal axis in Fig.\,\ref{fig5}(a1) and Figs.\,\ref{fig12}(a1)-\ref{fig12}(f1).}
\label{fig13}
\end{figure*}

As described before in Fig.\,\ref{fig7}, for understanding the self-healing, we have again analyzed the intensity cross-sections of propagated CADB with outer rings blocked (contains only $0^{\mathrm{th}}$ and $1^{\mathrm{st}}$ inner rings), as shown in Fig.\,\ref{fig13}. Figure\,\ref{fig13}(a) shows the normalized intensity cross-sections of an ideal CADB (blue solid curve) and partially blocked CADB (red dashed curve) at $z=0$, indicating that partially blocked CADB consists of only $0^{\mathrm{th}}$ and $1^{\mathrm{st}}$ inner rings. Figures\,\ref{fig13}(b)-\ref{fig13}(f) show the intensity cross-sections of an ideal CADB and partially blocked CADB at the propagation distances of $z=200$ mm, $400$ mm, $470$ mm, $550$ mm, and $750$ mm, respectively. Note, for the better visualization and comparison, in Figs.\,\ref{fig13}(b), \ref{fig13}(c) and \ref{fig13}(d), the intensity cross-section of partially blocked CADB is multiplied by the factors of 4, 40 and 2, respectively. As observed earlier in Fig.,\ref{fig7}, the blocked parts of the CADB again self-heals by redistribution of intensity during the propagation. This is evidenced by the appearance of multi-ring structure as the beam approach towards autofocusing distance, and shows more similarity to an ideal CADB. Note, the autofocusing distance of this partially blocked CADB and ideal CADB are 470 mm and 388 mm, respectively. This is also supported by an increase in the value of overlap integral, as marked by $C$ in Figs.\,\ref{fig13}(a)-\ref{fig13}(f).

\begin{figure}[htbp]
\centering   
\includegraphics[height = 7.5cm, keepaspectratio = true]{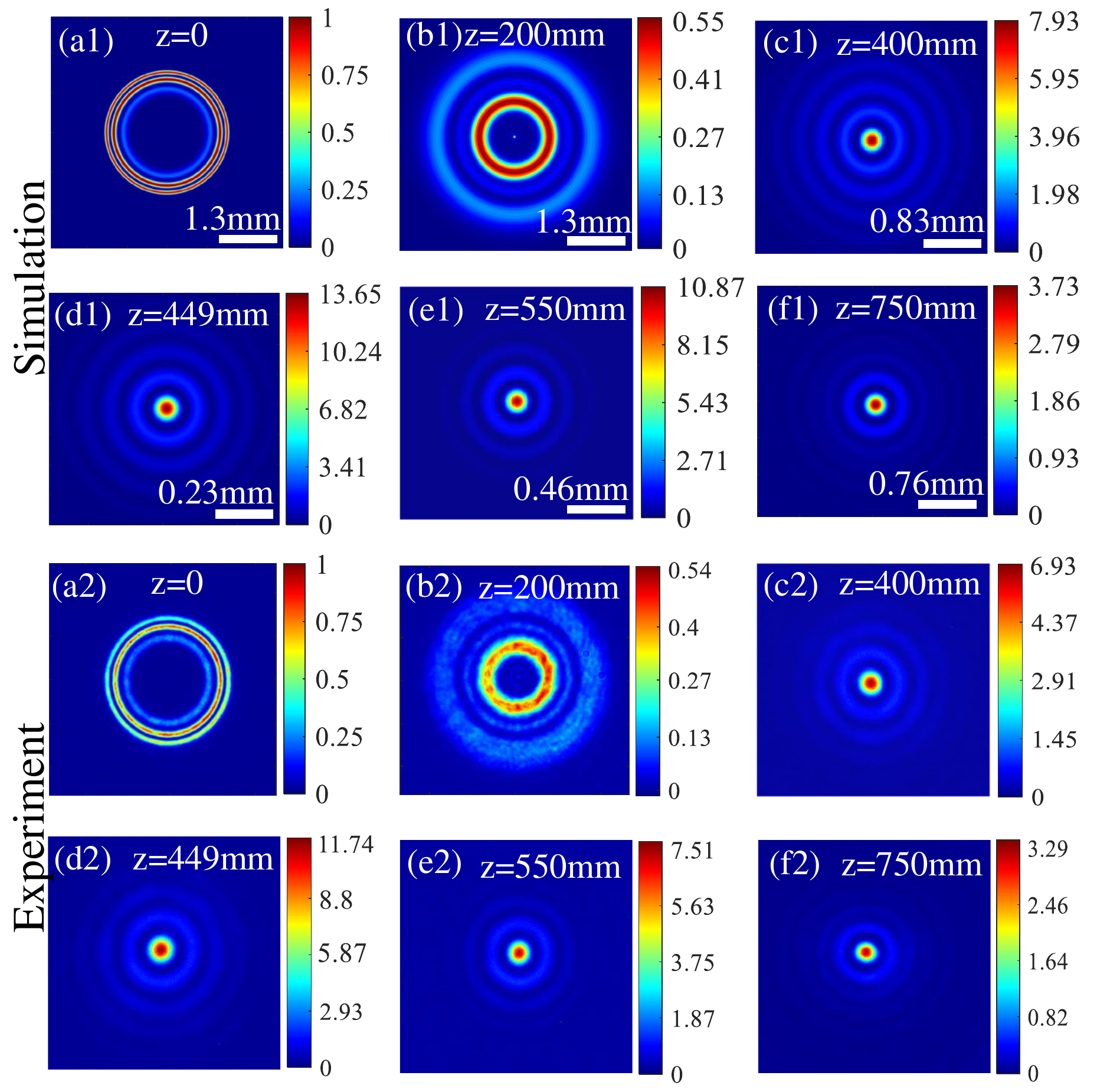}
\caption{Simulated and experimental intensity distributions of partially blocked CADB (contains only $0^{\mathrm{th}}$, $1^{\mathrm{st}}$ and $2^{\mathrm{nd}}$ inner rings) at different propagation distances (a1, a2) $z = 0$, (b1, b2) $z=200$ mm, (c1, c2) $z=400$ mm, (d1, d2) $z=449$ mm, (e1, e2) $z=550$ mm and (f1, f2) $z=750$ mm. The maximum value in colorbar represents the k-value. The CADB parameters are kept the same as in Fig.\,\ref{fig5}. Note, the scalebars in experimental results are the same at respective $z$.}
\label{fig14}
\end{figure}

Further, we have blocked the outer rings in CADB such that only $0^{\mathrm{th}}$, $1^{\mathrm{st}}$ and $2^{\mathrm{nd}}$ inner rings are left at $z=0$. This is obtained by using a binary annular aperture ( Eq.\,(\ref{eq3})) with the values of $r_m=0$ and $r_n=1.48$ mm. The results are shown in Fig.\,\ref{fig14}. Figures\,\ref{fig14}(a1) and \ref{fig14}(a2) show the simulated and experimental intensity distributions of blocked CADB, respectively, indicating that the outer rings are blocked and consist of only $0^{\mathrm{th}}$, $1^{\mathrm{st}}$ and $2^{\mathrm{nd}}$ inner rings. This leads to a reduced diffraction efficiency of $\eta=29.87\%$, indicating that a large amount of intensity ($70.13\%$) is blocked. Figures\,\ref{fig14}(b1)-\ref{fig14}(f1) and Figs.\,\ref{fig14}(b2)-\ref{fig14}(f2) show the simulated and experimental intensity distributions of partially blocked CADB propagated at different distances $z=200$ mm, $400$ mm, $449$ mm, $550$ mm, and $750$ mm, respectively. As evident, when partially blocked CADB is propagated, the intensity redistributes and moves into the blocked parts, and possesses the autofocusing at a distance of $z_{af}=449$ mm. This autofocusing distance is found to be smaller than that of partially blocked CADB having only $0^{\mathrm{th}}$ and $1^{\mathrm{st}}$ inner rings ($z=470$ mm) (Figs.\,\ref{fig12}(d1) and \ref{fig12}(d2)). Further, a similar trend in the k-value is observed. Particularly, the k-value increases with the propagation distance and reaches a maximum value at $z_{af}$, and after that it decreases. The maximum k-value at $z_{af}$ is found to be $\sim13.7$ in the simulation and $\sim11.7$ in the experiment, which are higher than that of partially blocked CADB having only $0^{\mathrm{th}}$ and $1^{\mathrm{st}}$ inner rings (Figs.\,\ref{fig12}(d1) and \ref{fig12}(d2)). However, these maximum k-value are found to be much smaller than that of an ideal CADB (Figs.\,\ref{fig5}(d1) and \ref{fig5}(d2)). The results clearly indicate that although the outer rings contain low intensities, but these play a significant role in autofocusing of CADB. 

\begin{figure}[htbp]
\centering
\includegraphics[height = 3.4cm, keepaspectratio = true]{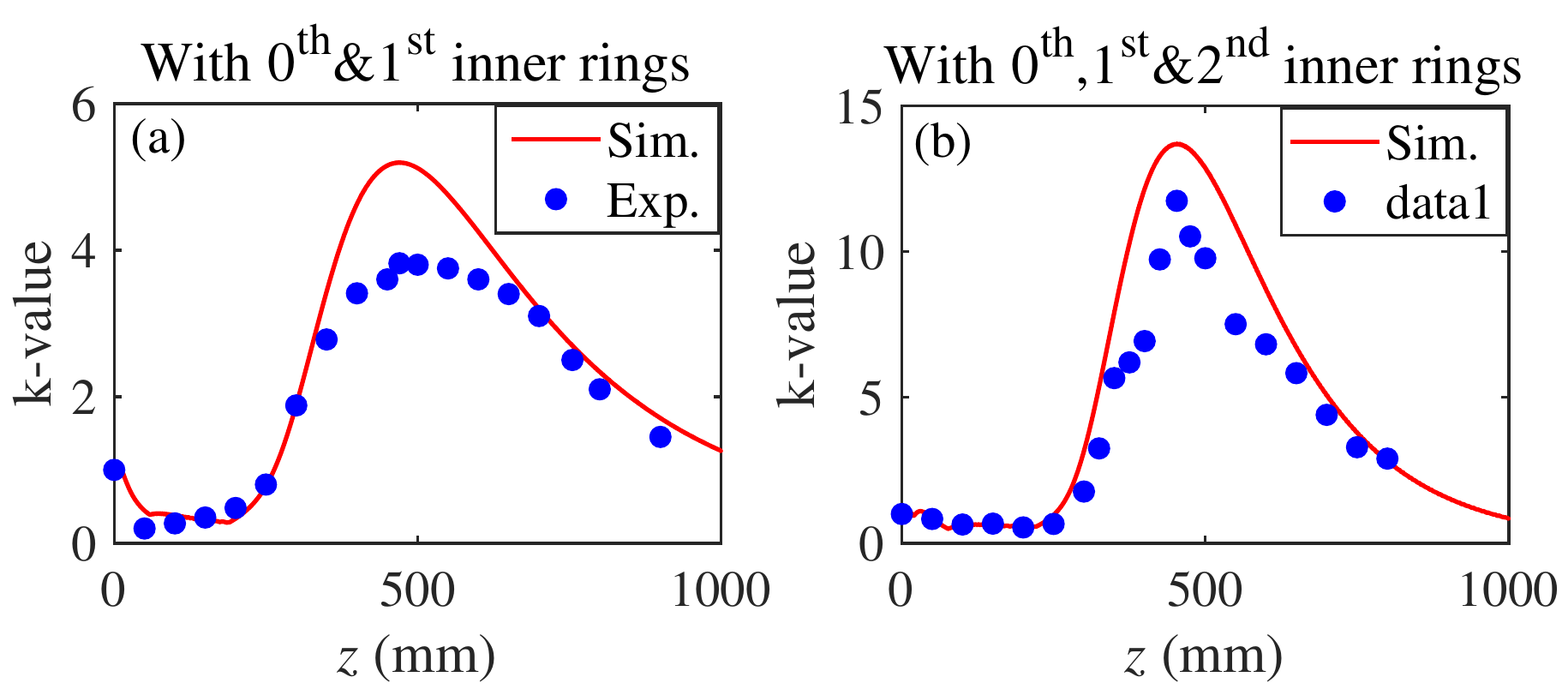}
\caption{Experimental (blue filled circles) and simulated (red solid curve) k-value as a function of propagation distance for blocked CADB having only (a) $0^{\mathrm{th}}$ and $1^{\mathrm{st}}$ inner rings, and (b) $0^{\mathrm{th}}$, $1^{\mathrm{st}}$ and $2^{\mathrm{nd}}$ inner rings.}
\label{fig15}
\end{figure}

A comparison of simulated (solid red curve) and experimental (blue filled circles) k-value as a function of propagation distance for the partially blocked CADBs (contains only $0^{\mathrm{th}}$ $\&$ $1^{\mathrm{st}}$ inner rings and $0^{\mathrm{th}}$, $1^{\mathrm{st}}$ $\&$ $2^{\mathrm{nd}}$ inner rings) are shown in Fig.\,\ref{fig15}. As evident, the k-value is found to be maximum at the autofocusing distances of $470$ mm (Fig.\,\ref{fig15}(a)) and $449$ mm (Fig.\,\ref{fig15}(b)).
Further, unlike ideal CADB and CADB blocked with inner rings (Fig.\,\ref{fig7}), in CADB blocked with outer rings the k-value gradually increases and decreases around the autofocusing position (large width of peak), which clearly suggests the lack of abrupt autofocusing. It indicates that the abrupt autofocusing characteristics of CADB strongly depends on low intensity outer rings.

\begin{figure}[htbp]
\centering
\includegraphics[height = 5cm, keepaspectratio = true]{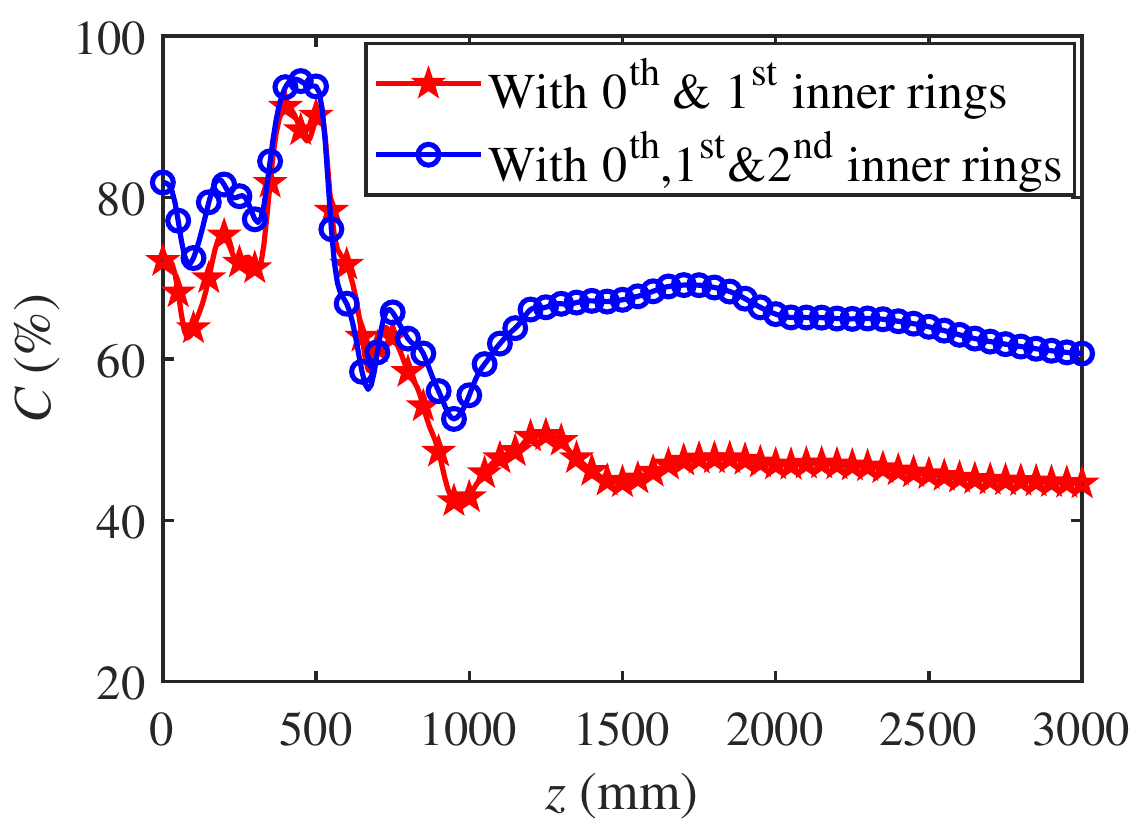}
\caption{The overlap integral as a function of propagation distance, for partially blocked CADB having only  $0^{\mathrm{th}}$ $\&$ $1^{\mathrm{st}}$ inner rings (red solid curve with stars) and partially blocked CADB having only $0^{\mathrm{th}}$, $1^{\mathrm{st}}$ $\&$ $2^{\mathrm{nd}}$ inner rings (blue solid curve with circles).}
\label{fig16}
\end{figure}

To quantify the self-healing, we have analyzed the overlap integral (Eq.\,(\ref{eq10})), as shown in Fig.\,\ref{fig16}. The red solid curve with stars denotes the results for partially blocked CADB having only $0^{\mathrm{th}}$ $\&$ $1^{\mathrm{st}}$ inner rings, whereas, blue solid curve with circles denote the results for partially blocked CADB having only $0^{\mathrm{th}}$, $1^{\mathrm{st}}$ $\&$ $2^{\mathrm{nd}}$ inner rings. At $z=0$, due to partial blocking of CADBs the reduced values of overlap $C$ are found to be $\sim72\%$ and  $\sim81\%$ corresponding to CADB having $0^{\mathrm{th}}$ $\&$ $1^{\mathrm{st}}$ inner rings and $0^{\mathrm{th}}$, $1^{\mathrm{st}}$ $\&$ $2^{\mathrm{nd}}$ inner rings, respectively. For $z>0$, the overlap initially fluctuates due to redistribution of intensity and reaches a maximum value at the autofocusing distance and after that it again decreases and shows fluctuations, and finally it attains a fixed value at large distances. In particular, for partially blocked CADB having only $0^{\mathrm{th}}$ $\&$ $1^{\mathrm{st}}$ inner rings, the overlap integral is found to be maximum $C=91\%$ at two points just before and after $z_{af}$ at $z=400$ mm and $z=510$ mm. However, at $z_{af}=470$ mm, the overlap is found to be $C=86\%$. At large distances the value of overlap is found to be $\sim44\%$. For partially blocked CADB having only $0^{\mathrm{th}}$, $1^{\mathrm{st}}$ $\&$ $2^{\mathrm{nd}}$ inner rings, the maximum value of overlap is found be $C=94\%$ at the autofocusing distance $z_{af}=449$ mm. At large distances the overlap $C$ is found to be $\sim60\%$. The results clearly indicate that the maximum self-healing of partially blocked CADBs (outer rings blocked) occurs either at $z_{af}$ or near to $z_{af}$. 

\subsection{Asymmetric blocking of CADBs} \label{asym}
\begin{figure}[htbp]
\centering
\includegraphics[height = 7.5cm, keepaspectratio = true]{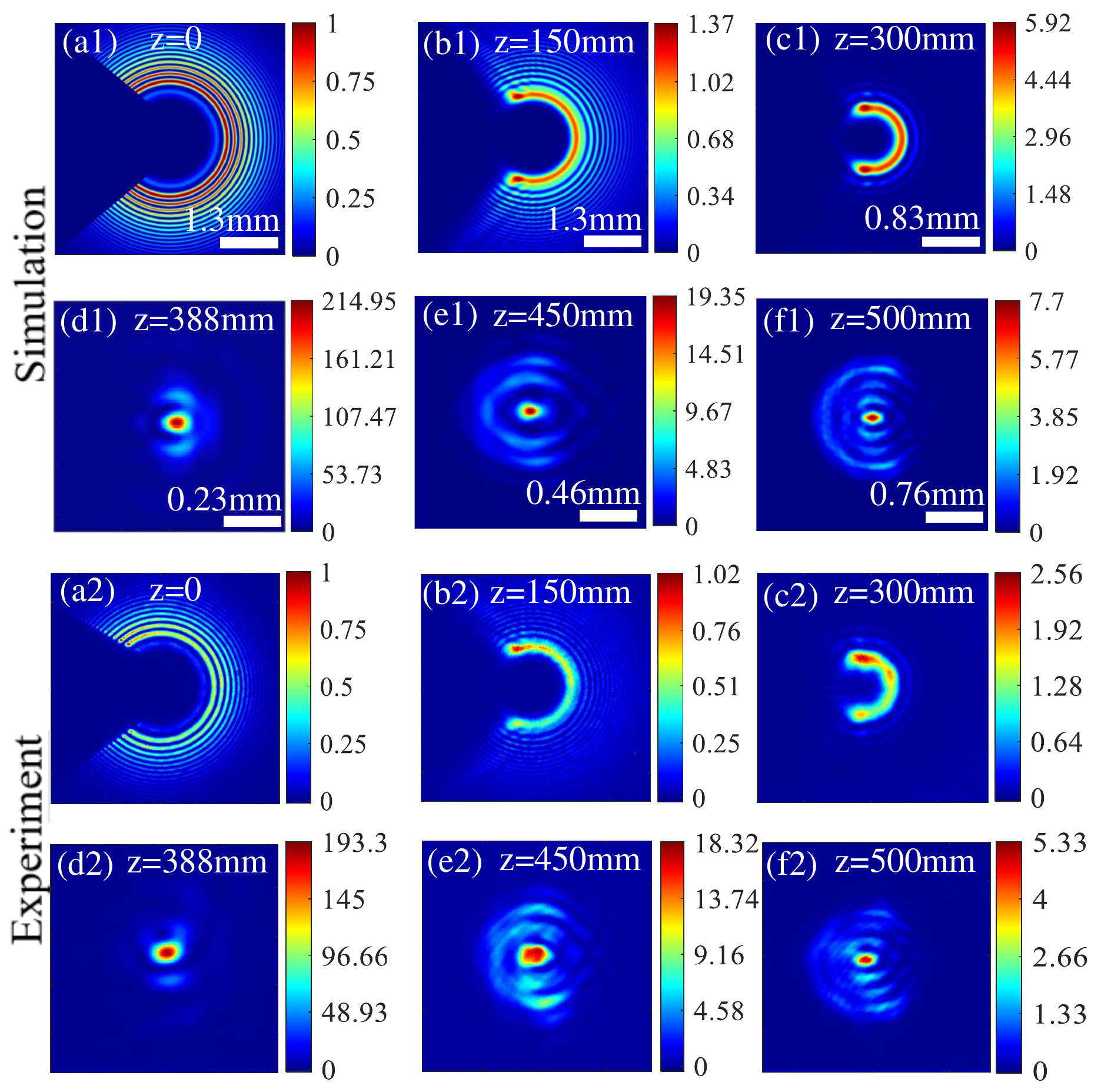}
\caption{Simulated and experimental intensity distributions of sectorially blocked CADB at different propagation distances (a1, a2) $z = 0$, (b1, b2) $z=150$ mm, (c1, c2) $z=300$ mm, (d1, d2) $z=388$ mm, (e1, e2) $z=450$ mm, and (f1, f2) $z=500$ mm. The maximum value in colorbar represents the k-value. The parameter values are kept the same as in Fig.\,\ref{fig5}. Note, the scalebars in experimental results are the same at respective $z$.} 
\label{fig17}
\end{figure}

Next, we have investigated the autofocusing and self-healing of CADB asymmetrically blocked with a sectorial aperture of $84^{\circ}$ (Figs.\,\ref{fig2}(c) and \ref{fig2}(f)) at $z=0$. The simulated and experimental results are shown in Fig.\,\ref{fig17}. Figures\,\ref{fig17}(a1) and \ref{fig17}(a2) show the simulated and experimental intensity distributions of sectorially blocked CADB at $z=0$, respectively. This sectorial blocking of CADB leads to a reduced diffraction efficiency of $\eta=76.68\%$, indicating that $23.32\%$ of total intensity is blocked by an aperture. Figures\,\ref{fig17}(b1)-\ref{fig17}(f1) and Figs.\,\ref{fig17}(b2)-\ref{fig17}(f2) show the simulated and experimental intensity distributions of partially blocked CADB at different propagation distances $z=150$ mm, $300$ mm, $388$ mm, $450$ mm, and $500$ mm, respectively. As evident, a sectorially blocked CADB possesses an autofocusing at a distance of $z_{af}=388$ mm (Figs.\,\ref{fig17}(d1) and \ref{fig17}(d2)), which is identical to the autofocusing distance of an ideal CADB (Figs.\,\ref{fig5}(d1) and \ref{fig5}(d2)). After the autofocusing, the intensity again redistributes and forms multiple rings with retaining a bright spot at the center (Figs.\,\ref{fig17}(e1)-\ref{fig17}(f1) and Figs.\,\ref{fig17}(e2)-\ref{fig17}(f2)). The k-value increases with the distance and reaches a maximum value at $z_{af}$, and after that it decreases. The maximum k-value at $z_{af}$ is found to be $\sim214.95$ in the simulation and $\sim 193$ in the experiment. These values are found to be smaller than that of an ideal CADB (Fig.\,\ref{fig5}).

\begin{figure}[htbp]
\centering
\includegraphics[height = 3.0cm, keepaspectratio = true]{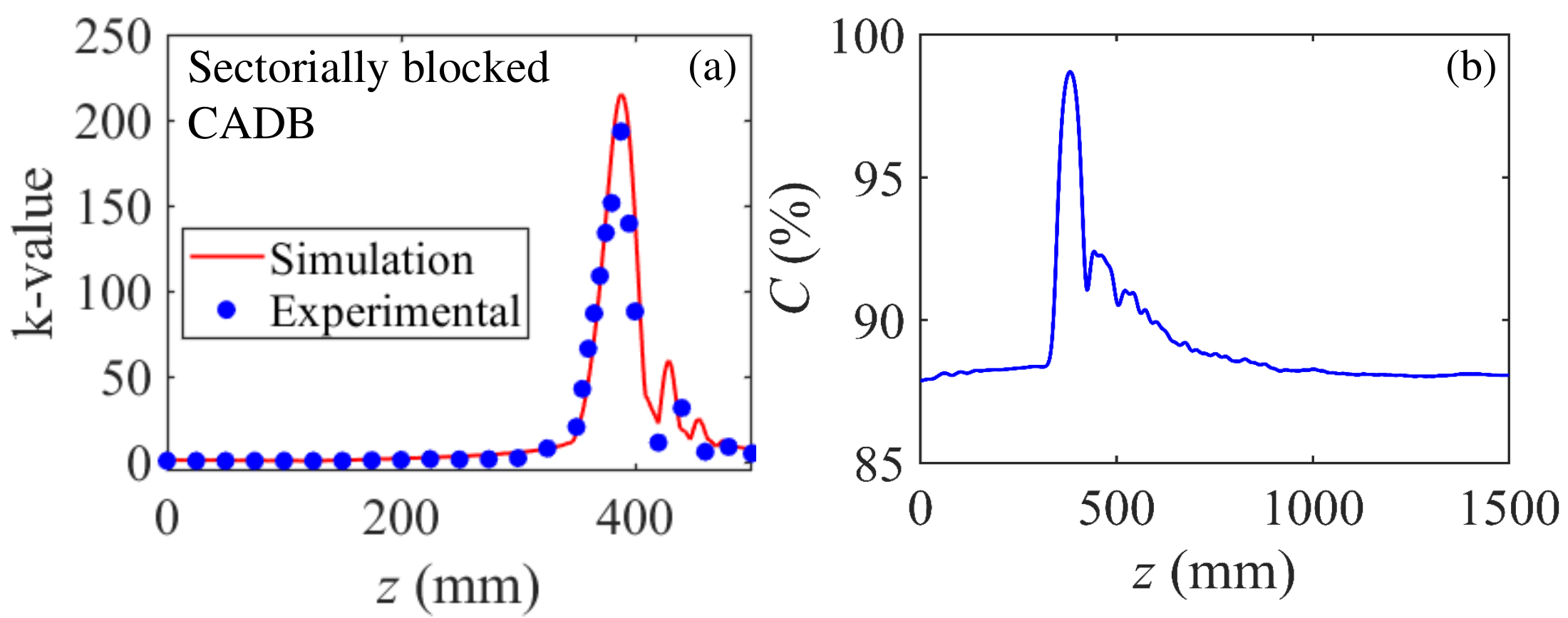}
\caption{(a) Simulated (red solid curve) and experimental (blue filled circles) k-value of sectorially blocked CADB as a function of propagation distance. (b) The overlap integral ($C$) of sectorially blocked CADB as a function of propagation distance.}
\label{fig18}
\end{figure}

 The simulated (red solid curve) and experimental (blue filled circles) k-value as a function of propagation distance for a sectorially blocked CADB are shown in Fig.\,\ref{fig18}(a). As evident, the k-value shows a sharp variation around the autofocusing distance in the form of a narrow peak, which indicates an abrupt autofocusing characteristics of sectorially blocked CADB. The results show a good agreement between the simulation and experiment.

To quantify the self-healing of sectorially blocked CADB, we have analyzed the overlap integral (Eq.\,\ref{eq10}) with respect to an ideal CADB having the same parameters, as shown in Fig.\,\ref{fig18}(b). As evident, at $z=0$ due to sectorial blocking of CADB, the value of overlap $C$ is reduced to $\sim88\%$. The overlap $C$ remains almost unchanged up to a distance of $z<317$, and then it increases sharply at the autofocusing distance ($z_{af}=388$ mm) and attains a maximum value of $\sim98\%$. After $z_{af}$, the overlap decreases and shows some fluctuations due to redistribution of intensity, and finally attains a constant value of $\sim 88\%$. The results clearly evidence that a sectorially blocked CADB self-heals maximally by propagation up to the autofocusing distance.  
 
\section{Comparison of partially blocked CADB and ideal CADB having the same dark central region}
The abrupt autofocusing features of CADB can be controlled by the beam parameters, such as radius $r_{0}$, exponential decay factor $a$, and scaling factor $w_{0}$ (Eq.\,(\ref{eq1})) \cite{zang2022}]. It has been shown that by increasing the radius $r_{0}$, the k-value enhances and the autofocusing distance increases \cite{zang2022}. The smaller truncation factor provides stronger autofocusing abilities (increased k-value), however, the autofocusing distance remains nearly unchanged. Further, a decrease in scaling factor $w_{0}$ results in a decrease in the autofocusing distance, however, the abrupt autofocusing ability enhances \cite{zang2022}.
\begin{figure}[htbp]
\centering
\includegraphics[height = 7.5cm, keepaspectratio = true]{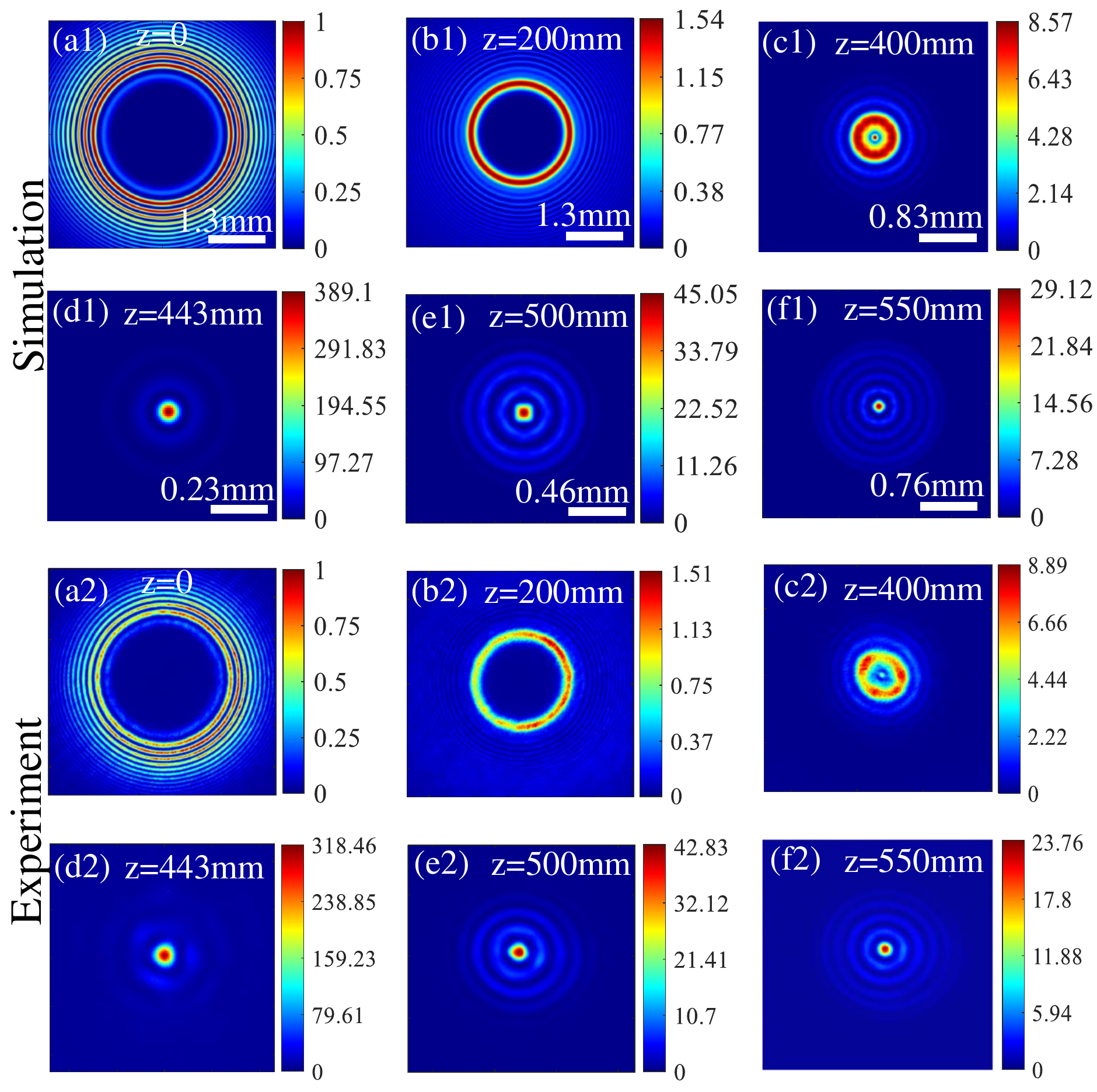}
\caption{Simulated and experimental intensity distributions of ideal CADB with increased radius at different propagation distances (a1, a2) $z = 0$, (b1, b2) $z=200$ mm, (c1, c2) $z=400$ mm, (d1, d2) $z=443$ mm, (e1, e2) $z=500$ mm, and (f1, f2) $z=550$ mm. The parameter values are taken as $r_0=1.32$ mm, $w_0=0.1$ mm, $a=0.1$ and $\lambda=1064$ nm. The maximum value in colorbar represents the k-value. Note, the scalebars in experimental results are the same at respective $z$.}
\label{fig19}
\end{figure} 

We have compared the abrupt autofocusing properties of partially blocked CADB and ideal CADB having the same radius $r_0$ at $z=0$. Our aim is to see whether these CADBs having the same dark central regions possess the same or different behaviours. A partially blocked CADB is obtained by blocking from the center ($r=0$) up to first inner ring of an ideal CADB (Fig.\,\ref{fig5}), and its radius becomes $r_{0}=1.32$ mm. The results of this partially blocked CADB are shown in Fig.\,\ref{fig6}. For the comparison, we have considered an ideal CADB with the same radius of $r_{0}=1.32$ mm. The simulated and experimental results of propagation of an ideal CADB with this increased radius are shown in Fig.\,\ref{fig19}. Figures\,\ref{fig19}(a1)-\ref{fig19}(f1) and Figures\,\ref{fig19}(a2)-\ref{fig19}(f2) show the simulated and experimental intensity distributions of an ideal CADB with increased radius at different propagation distances, respectively. As evident, the beam shows the autofocusing at $z_{af}=443$ mm (Figs.\,\ref{fig19}(d1) and \ref{fig19}(d2)), and intensity sharply focused in the form of a bright central spot. After $z_{af}$, the intensity again redistributes in the form of multiple rings with retaining a bright central spot. The k-value follows a similar trend as observed earlier, where it slowly increases in the beginning, and as the propagation distance approaches towards autofocusing it sharply increases to a maximum value, and after that rapidly decreases, which confirms the abrupt autofocusing behaviour. Unlike an ideal CADB with radius $r_{0}=1$ mm (Fig.\,\ref{fig5}), an ideal CADB with increased radius $r_{0}=1.32$ mm possesses abrupt autofocusing at an increased distance, agrees with the findings of \cite{zang2022}. The k-value for an ideal CADB with increased radius is found to be $\sim 389$ in the simulation and $\sim 318$ in the experiment, which are higher than that of an ideal CADB with $r_{0}=1$ mm (Fig.\,\ref{fig5}), which again agrees with the findings of \cite{zang2022}.

\begin{figure}[htbp]
\centering
\includegraphics[height = 5cm, keepaspectratio = true]{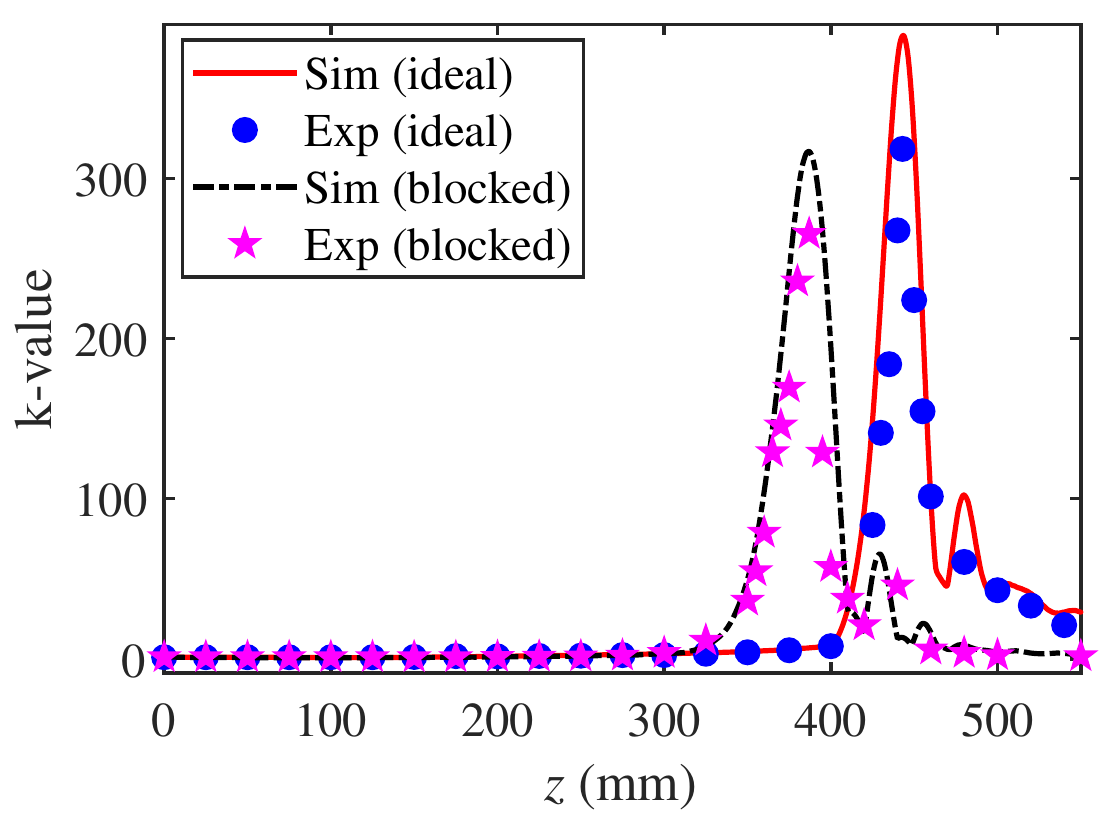}
\caption{Experimental and simulated k-value for an ideal CADB with increased radius ($r_0$ = 1.32 mm) (red solid curve and blue filled circles) and partially blocked CADB having the same radius $r_{0}=1.32$ mm (dot-dashed black curve and pink stars).}
\label{fig20}
\end{figure}

Further, an ideal CADB with increased radius ($r_{0}=1.32$ mm) autofocuses at a longer distance as compared to the partially blocked CADB with the same radius ($z_{af}=387$ mm) (Figs.\,\ref{fig6}(d1) and \ref{fig6}(d2)). For an ideal CADB and partially blocked CADB with the same radii, the k-value as a function of propagation distance is shown in Fig.\,\ref{fig20}. As evident, the maximum k-value occurs at different autofocusing distances ($z_{af}=443$ mm for ideal CADB and $z_{af}=387$ mm for partially blocked CADB). Further, the k-value sharply varies around the autofocusing distances and exhibits narrow peaks, which confirms the abrupt autofocusing behavior of both ideal CADB and partially blocked CADB having the same radius. In an ideal CADB the maximum k-value at $z_{af}$ is found to be larger than that of a partially blocked CADB with the same radius. This suggests that an ideal CADB possesses better abrupt autofocusing features. However, in common CAB the opposite behavior is observed, where k-value increases more in blocked CAB than that of an ideal CAB with the same radius \cite{li2014}. 
\section{Conclusions}\label{concl}
In conclusion, we have investigated autofocusing and self-healing properties of CADBs blocked symmetrically and asymmetrically in an initial plane $z=0$. In symmetric blocking, we have truncated inner and outer rings of CADB, whereas, in asymmetric blocking, CADB is truncated sectorially. We have found that when CADB is blocked with first inner ring, two inner rings, and three inner rings, it still possesses strong abrupt autofocusing and the autofocusing distance ($z_{af}$) remains unchanged. The $z_{af}$ is found to be similar to that of an ideal CADB having the same beam parameters. In all three blocked cases of inner rings, the k-value shows a similar trend with respect to the propagation distance $z$, and in each case, the maximum k-value occurs at $z_{af}$. However, the maximum k-value is found to be decreased with the number of inner rings blocked. For example, it is found to be largest in the case of first inner ring blocked. The k-value shows a sharp variation (increase and decrease) around $z_{af}$, which suggests that CADBs blocked with inner rings possess abrupt autofocusing characteristics. Further, CADBs blocked with inner rings also possess good self-healing abilities, which occurs due to the redistribution of intensity into blocked parts during the propagation. In all three cases of blocked inner rings, the maximum self-healing occurs at $z_{af}$, which is quantified by an overlap integral $C$.

For CADBs blocked with outer rings, $z_{af}$ changes with the number of outer rings blocked. In particular, for blocking more number of outer rings the $z_{af}$ increases. The k-value does not show an abrupt variation around $z_{af}$, which shows an absence of abrupt autofocusing features. The k-value is found to be much smaller than that of ideal CADB, however, its maximum value still occurs at $z_{af}$. The maximum k-value again decreases with the increase in the number of blocked outer rings. The CADB blocked with outer rings still possesses good self-healing abilities, and maximum self-healing occurs either at $z_{af}$ or very near to $z_{af}$. Although, the outer rings consist of low intensities, but these play a significant role in autofocusing and self-healing of CADB.

In asymmetric blocking, a sectorially blocked CADB possesses good abrupt autofocusing and self-healing properties. The $z_{af}$ is found to be the same as for an ideal CADB ($z_{af}=388$ mm). However, the maximum k-value at $z_{af}$ is found to be reduced as compared to an ideal CADB. The maximum self-healing is again found to be at $z_{af}$.

Further, we have compared the autofocusing features of an ideal CADB and a partially blocked CADB having the same radii (with the same amount of dark central regions). The ideal CADB shows more pronounced abrupt autofocusing (k-value found to be large), however, the autofocusing distance becomes longer. These comparative results are found to be different than a common CAB \cite{li2014}. We have found a reasonably good agreement between the numerical simulations and experimental results.

Due to good autofocusing and self-healing abilities of partially blocked CADBs, these beams are beneficial for applications in many fields  such as in biomedical treatment where the light travels in disordered media, optical manipulation and trapping and also for imaging where deep penetration with small and stronger focus is required. 

\section*{Acknowledgment}
We acknowledge the funding support from Indian Institute of Technology Ropar (Grant no.: 9-230/2018/IITRPR/3255) and Science and Engineering Research Board (Grant no. CRG/2021/003060).  Anita Kumari and Vasu Dev acknowledge the fellowship support from Unversity Grants Commission (UGC) and IIT Ropar.

%\bibliography{Bibliography}

\end{document}